\documentclass[aps,pre,reprint,groupedaddress]{revtex4-2}

\usepackage{graphicx}   
\usepackage{dcolumn}    
\usepackage{bm}         
\usepackage{amsmath}    
\usepackage{amssymb}    
\usepackage{mathtools}

\begin{document}

\title{Cross Validation in Stochastic Analytic Continuation}

\author{Gabe Schumm}
\email{gschumm@bu.edu}
\affiliation{Department of Physics, Boston University, 590 Commonwealth Avenue, Boston, Massachusetts 02215, USA}

\author{Sibin Yang}
\affiliation{Department of Physics, Boston University, 590 Commonwealth Avenue, Boston, Massachusetts 02215, USA}

\author{Anders W. Sandvik}
\email{sandvik@bu.edu}
\affiliation{Department of Physics, Boston University, 590 Commonwealth Avenue, Boston, Massachusetts 02215, USA}

\date{\today}

\begin{abstract}
 Stochastic Analytic Continuation (SAC) of Quantum Monte Carlo (QMC) imaginary-time correlation function data is a valuable tool in connecting many-body models to experimentally measurable dynamic response functions. Recent developments of the SAC method have allowed for spectral functions with sharp features, e.g. narrow peaks and divergent edges, to be resolved with unprecedented fidelity. Often times, it is not known what exact sharp features, if any, are present \textit{a priori}, and, due to the ill-posed nature of the analytic continuation problem, multiple spectral representations may be acceptable. In this work, we borrow from the machine learning and statistics literature and implement a cross validation technique to provide an unbiased method to identify the most likely spectrum amongst a set obtained with different spectral parameterizations and imposed constraints. We demonstrate the power of this method with examples using imaginary-time data generated by QMC simulations and synthetic data generated from artificial spectra. Our procedure, which can be considered a form of model selection, can be applied to a variety of numerical analytic continuation methods, beyond just SAC.
\end{abstract}

\maketitle

\section{Introduction}
Dynamic response functions (spectral functions) provide a key link between the quantum Monte Carlo (QMC) simulations of many-body systems and their experimental counterparts. While it is intractable to measure these spectral function directly by QMC simulations, one can instead perform numerical analytic continuation of imaginary-time correlation functions to real frequency. However, this process is what is considered an ``ill-posed'' problem \cite{schuttler_85, white_89,  jarrell_89}, meaning that there may be many viable solutions for the spectral function fitting a set of QMC-generated data. The Maximum Entropy method (MEM) \cite{silver_90_1, silver_90_2, gubernatis_91,gubernatis_96} has been the traditional tool used to tackle this problem and has produced many important results \cite{bergeron_16}. However, MEM has its limitations, being able only to reproduce broad spectral features, unless prior information about more complicated structure is provided to the algorithm. 

An alternative approach to MEM is to instead rely on the stochastic averaging of many solutions for the spectral function that fit the QMC-generated data well. The Stochastic Analytic Continuation (SAC) method \cite{white_91, sandvik_98, beach_04, vafayi_07, reichman_09, olav_08, fuchs_10, sandvik_16, qin_17, ghanem_20,  koch_20, ghanem_20, shao_23, ghanem_23}, also referred to as the average spectrum method, achieves this by regularizing the average spectra using a fictitious temperature, $\Theta$, which mediates the competition between the minimization of the goodness-of-fit $\chi^2$ with respect to the QMC-generated data and the favoring of spectra with the high configurational entropy. It is now known that this procedure is equivalent to the MEM when the number of sampled degrees of freedom of the spectrum is large and when the two resulting spectra are compared at the same $\chi^2$ value (although the definition of the entropy must be adjusted depending on what degrees of freedom are sampled) \cite{beach_04, shao_23, ghanem_23}. Recent developments of the SAC method have also allowed for the resolution of spectra with sharp features, such as narrow quasi-particle peaks and edges with power-law singularities, by imposing appropriate constraints on the sampling space of the spectral function \cite{sandvik_16, shao_23}.

In both the plain, ``unconstrained'' version of SAC, and the newly developed ``constrained'' SAC sampling schemes, the average $\chi^2$ of the sampled spectra is used to optimize various parameters and gauge the ability of the spectral representation to describe the data. While these methods that rely on the average $\chi^2$ alone have been shown to produce reliable and consistent results, the goal of this study is to provide further predictive power for these simple schemes by taking advantage of concepts from the machine learning and statistics literature. We do this by implementing a cross validation procedure \cite{bishop_06, mehta_19}. 

Cross validation is a tool used to compare how well different models describe a set of data. This is generally achieved by splitting the available data into mutually exclusive sets; a training, or sampling, set and a series of testing, or validation, sets. A model is trained on the sampling set, typically by optimizing some type of loss function. After training, the ability of the optimized model to describe the validation data sets is determined using this same loss function. This can provide information about the suitability of the model to describe the data, as well as the level of over-fitting to noise in the sampling data set. 

Cross validation has a natural extension to SAC. The imaginary-time correlation function data, calculated using QMC simulations, are typically stored as individual bins, which can be split up into sampling and validation sets. Here, the goodness-of-fit $\chi^2$ is used as the loss function to ``train'' and to ``validate'' the resulting spectral function. The concept of using cross validation within SAC was first explored by Efremkin et al.~\cite{efremkin_21}, whose work we draw on and significantly expand upon. Here, we perform an exhaustive study using both synthetic and real QMC-generated data. 

In Sec.~\ref{procedure} we describe the SAC procedure generally, and in Sec.~\ref{cross_val_proc} we explain our cross validation scheme in detail. In Sec.~\ref{temp_A} and ~\ref{temp_B} we implement cross validation using the unconstrained SAC method and demonstrate how it can be used to determine the optimal sampling temperature, which we compare to the simple criterion proposed previously \cite{shao_23}. In Sec.~\ref{double_peak} and ~\ref{shoulder}, we show how cross validation can be used in practice to determine which \textit{unconstrained} SAC sampling parameterization produces the most statistically likely spectrum. We extend this model selection procedure to the class of \textit{constrained} SAC parameterization in Sec.~\ref{constrained}. In Sec.~\ref{spec_param} we perform a test using QMC-generated data for the $S=1/2$ antiferromagnetic (AFM) Heisenberg model, where it is known that the spectral function for the operator $O=S_q^z$ contains a power-law divergent edge. In Sec.~\ref{spec_param_2} we perform a preliminary test on the  $S=1/2$ AFM Heisenberg model with long-range interactions \cite{feiguin_21, yang_24}, a model where the exact features of the spectral function are unknown. We summarize our results and discuss the remaining open questions and future directions in Sec.~\ref{conc}.

\section{SAC Procedure}\label{procedure}
For the purpose of self-containment and to set the notation, we will first provide a brief overview of the SAC procedure, as well as the small adjustments needed for cross validation.

 We consider the imaginary-time correlation function of an operator $O$ at inverse temperature $1/T =\beta$, evaluated at time points $\tau \in \left[0, \beta/2\right]$,
\begin{equation}
G(\tau) = \langle O^\dagger (\tau) O(0)\rangle.
\end{equation}
Here we consider only bosonic operators, but this procedure can be extended to fermionic operators as well, with only minor adjustments to the SAC procedure. The imaginary-time dependence is defined in the Heisenberg picture, taking $\hbar = 1$,
\begin{equation}
O (\tau) = e^{\tau H}O e^{-\tau H}.
\end{equation}
The corresponding spectral function can be defined in terms of the eigenstates $|n\rangle$ and energy eigenvalues $E_n$ of the Hamiltonian describing the system,
\begin{equation}\label{S_omega_eq}
S(\omega) = \frac{\pi}{Z} \sum_{m,n} e^{-\beta E_n} |\langle m|O|n\rangle|^2\delta\left(\omega - [E_m-E_n]\right),
\end{equation}
 where $Z$ is the partition function. The relationship between $S(\omega)$ and $G(\tau)$ is given by the equation
 \begin{equation}\label{G_tau}
G(\tau) = \int_0^\infty d\omega\,S(\omega) K(\tau, \omega),
\end{equation}
where the bosonic kernel $K(\tau, \omega)$ is defined as
\begin{equation}
K(\tau, \omega) = \frac{1}{\pi}\left(e^{-\tau \omega} + e^{-(\beta - \tau) \omega}\right).
\end{equation}

For the cross validation procedure, we consider $N_B$ bins of imaginary-time correlation function data evaluated at a set of $N_\tau$ points $\tau_i$, $\{G^b(\tau_i) \}$, where $b = 1, 2, \ldots, N_B$ and $i = 1, 2, \ldots, N_\tau$. The $N_B$ bins are then split into $K+1$ mutually exclusive sets $\{G^b(\tau_i) \}_k$ of equal length  $N_k= {N_B}/\left({K+1}\right)$. For a single cross validation run, one of the $K+1$ groups will act as the sampling data set, while the other $K$ groups will be used for cross validation. For each group, we compute the average
\begin{equation}
\bar{G}_{k}(\tau_i) = \frac{1}{N_k} \sum_{b \in k} G^b(\tau_i),
\end{equation}
where the sum is over all bins $b$ in the $k$th validation data set. We define the ``error level'' for these sets of $G(\tau)$ data as the magnitude of the error bar on the averaged correlation function, $\bar{G}_k(\tau_i)$, for the largest $\tau$ point included, when normalized so that $\bar{G}_k(0) = G^b(0) = 1$. For larger $\beta$, a cutoff is typically introduced at the $\tau$ value where the relative error on $\bar{G}_{k}(\tau) $ is greater than 10\%, which we would instead use to quantify the error level.

The error level provides a rough gauge of the data quality, but because the statistical errors of different $\tau$ points are correlated, the full characterization of the error, and thus the calculation of $\chi^2$, requires the covariance matrix \cite{gubernatis_96}. We use the bootstrap method to calculate the covariance matrices for each set, 
\begin{equation}
C_{k,ij} = \frac{1}{M} \sum_{m=1}^M (G_k^m(\tau_i) - \bar{G}_{k}(\tau_i)) (G_k^m(\tau_j) - \bar{G}_{k}(\tau_j) )
\end{equation}
where $G_k^m(\tau_i)$ is the average of a bootstrap sample of $N_k$ randomly chosen bins among the $N_k$ bins in the $k$th data set and $M$ is the total number of bootstrap samples, typically much larger than $N_B$. 

Given $S(\omega)$, the corresponding $G(\tau_i)$ values are computed according to Eq.~\eqref{G_tau} and the $\chi^2$ with respect to the QMC-generated data is given by   
\begin{align}
\chi^2_k = \sum_{i, j}^{N_\tau}[G(\tau_i) - \bar{G}_{k}(\tau_i)]\left[C_k^{-1}\right]_{ij}[G(\tau_j) - \bar{G}_{k}(\tau_j)].
\end{align}
Alternatively, one can compute $\chi^2_k$ in the eigenbasis of the covariance matrix: 
\begin{equation}
\chi^2_k = \sum_{i=1}^{N_\tau}\left(\frac{G(\tau_i) - \bar{G}_{k}(\tau_i)}{\sigma_k(\tau_i)}\right)^2,
\end{equation}
where $\left(\sigma_k(\tau_i)\right)^2$ are the eigenvalues of the covariance matrix and both $G(\tau_i)$ and $ \bar{G}_{k}$ have now been transformed to this eigenbasis.

\begin{figure}[t]
\begin{center}
\includegraphics[width=.94\columnwidth]{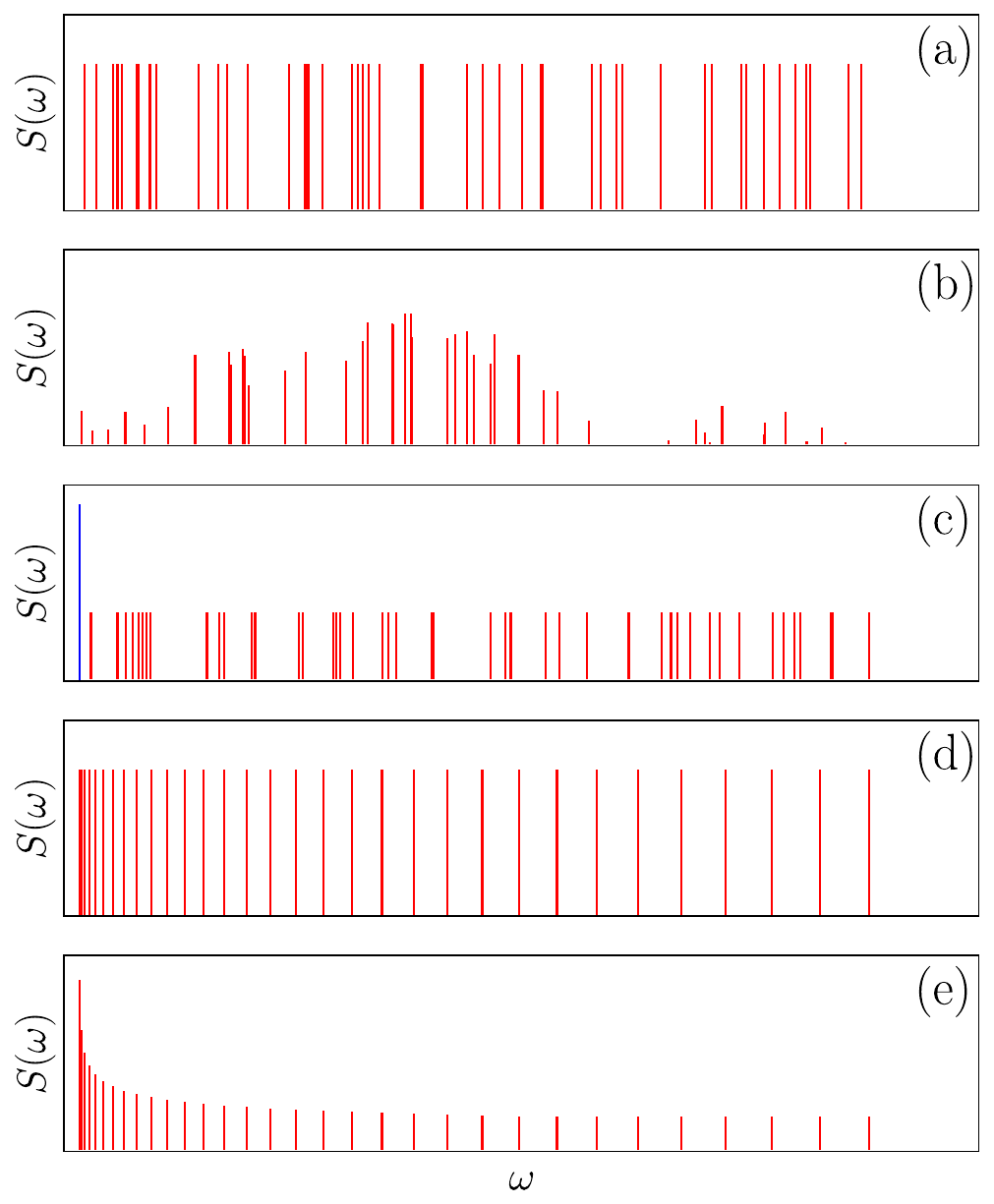}
\caption{Depictions of the parameterizations of the spectrum $S(\omega)$: (a) Equal amplitudes and sampled frequencies in the continuum. (b) Sampled amplitudes and sampled frequencies in the continuum. (c)  A ``macroscopic'' $\delta$-function with amplitude $A_0$ at $\omega_0$, followed by a large number $N_\omega$ of ``continuum'' $\delta$-functions at positions $\omega_i>\omega_0$ with equal amplitudes $A_i = (1-A_0)/N_{\omega}$. The amplitude $A_0$ is fixed in a sampling run, but the frequency $\omega_0$ is sampled. (d) Equal amplitudes with monotonically increasing spacing $d_i = \omega_{i+1} - \omega_i$. Just as in (b), the lowest frequency $\omega_1$ is sampled along with all other frequencies, with the constraint $d_{i+1} > d_i$. (e) Varying amplitudes with monotonically increasing spacing, as in (c). In all cases, the final spectrum is the mean amplitude density of the $\delta$-functions, which is accumulated in a histogram during the sampling process.}
\label{params}
\end{center}
\end{figure}

In a SAC run, the spectrum $S(\omega)$ is parameterized as a large number of $\delta$-functions in a frequency continuum (in practice a very dense frequency grid). As will be described in detail later, the amplitudes and positions of the collection of  $\delta$-functions can be subject to constraints to produce different spectral forms. The parameterizations (i.e. constraints) considered here are depicted in Fig.~\ref{params}.

The configurations of the $\delta$-functions are importance sampled with a weight 
\begin{equation}
P(S) \propto \exp\left( -\frac{\chi^2(S) }{2\Theta} \right).
\end{equation}
The fictitious temperature $\Theta$ is gradually reduced, as in simulated annealing, until a minimum is reached. In the limit $\Theta \to 0$, the sampled spectrum purely minimizes $\chi^2$, which will not reproduce the most probable spectrum when noisy data are used. Purely Bayesian arguments imply that the temperature should be fixed to $\Theta = 1$ \cite{white_91, olav_08}. However, for $S(\omega)$ parameterized by  a large number of sampled degrees of freedom, entropy will ruin the goodness-of-fit at this fixed value of $\Theta$ \cite{shao_23}. Instead, it has been proposed to use a temperature determined by a simple criterion to give the optimal balance of entropy and goodness-of-fit, corresponding to the value where 
\begin{equation}\label{criterion}
\langle \chi^2(\Theta)\rangle = \chi^2_{\rm min} + a \sigma_{\chi^2},
\end{equation}
where $\sigma_{\chi^2} = \sqrt{2\chi^2_{\rm min} }$ is used as a proxy for the standard deviation of the $\chi^2$ distribution that the value of $\langle \chi^2\rangle$ follows and $a$ is an order-one number (typically $a=1/2$ is used). This criterion, motivated by the properties of the $\chi^2$ distribution, has been shown to produce reliable results, reproducing spectra generated from synthetic QMC $G(\tau)$ data with high fidelity \cite{shao_23}.

\section{Cross Validation Procedure}\label{cross_val_proc}
We will now describe the implementation of cross validation within the SAC framework laid out above. Using the $ k = 0$th set of bins as the sampling data set, we carry out a simulated annealing run, as described above. Along with the sampling $\chi^2$, at each temperature we use the sets $k=1,\dots, K$ to  compute the validation $\chi^2$ value:
\begin{equation}\label{chi2_val}
\chi^2_{\mathrm{val}} = \frac{1}{K} \sum_{k=1}^K \sum_{i=1}^{N_\tau} \left( \frac{G(\tau_i) - \bar{G}_k(\tau_i)} {\sigma_k(\tau_i)}\right)^2.
\end{equation}

If we denote the exact correlation function by $G_{\mathrm{ex}}(\tau)$, Eq.~\eqref{chi2_val} can be rewritten as 
\begin{equation}
\chi^2_{\mathrm{val}} = \frac{1}{K} \sum_{k=1}^K \sum_{i=1}^{N_\tau} \left( \frac{G(\tau_i) -G_{\mathrm{ex}}(\tau_i) + G_{\mathrm{ex}}(\tau_i) - \bar{G}_k(\tau_i)} {\sigma_k(\tau_i)}\right)^2,
\end{equation}
which can be split into three terms:
\begin{align}\label{chi2_val_factor}
 &\chi^2_{\mathrm{val}}  =X_1 + X_2 + X_3,
\end{align}
where
\begin{subequations}\label{chi2_val_factor_abc}
    \begin{align}
    &X_1 = \frac{1}{K} \sum_{k=1}^K \sum_{i=1}^{N_\tau} \left( \frac{G(\tau_i) - {G}_{\rm ex}(\tau_i)} {\sigma_k(\tau_i)}\right)^2,\\
    &X_2 = \frac{1}{K} \sum_{k=1}^K \sum_{i=1}^{N_\tau} \left( \frac{\bar{G}_k(\tau_i) - {G}_{\rm ex}(\tau_i)} {\sigma_k(\tau_i)}\right)^2,\\
    &X_3 = \frac{2}{K} \sum_{k=1}^K \sum_{i=1}^{N_\tau} \frac{\left(\bar{G}_k(\tau_i) - G_{\rm ex}(\tau_i)\right) \left({G}_{\rm ex}(\tau_i) - {G}(\tau_i)\right) } {\sigma^2_k(\tau_i)}.
    \end{align}
\end{subequations}

Here $X_1$ contains purely information about how well the estimated $G(\tau)$, calculated from the sampled $S(\omega)$, describes the exact correlation function. This will vary as we change $\Theta$, or any other parameter used in the SAC program, and will be minimized by the most statistically likely spectral function, given input data $\bar{G}_0(\tau)$ and the corresponding covariance matrix $C_0$.

$X_2$ is the the $\chi^2$ value for the validation data with respect to the underlying data $G_{\rm ex}$, averaged over each of the $K$ validation data sets. The value of $X_2$ will follow the $\chi^2$ distribution with $N_{\rm dof}=N_\tau$ and will have a mean of $E\left[X_2\right] = N_{\rm dof} = N_\tau$ and a variance of  $V\left[X_2\right] = 2N_{\rm dof} = 2N_\tau$. This term is independent of any SAC parameter and will simply contribute a background value of $N_\tau$, in the limit of large $K$.

$X_3$ contains both information about the accuracy of $G(\tau)$ with respect to ${G}_{\rm ex}(\tau)$ and the deviation of the $\bar{G}_k(\tau)$ with respect to ${G}_{\rm ex}(\tau)$. However, these two factors are uncorrelated, and since we expect that the fluctuations of $\bar{G}_k(\tau)$ will follow a Gaussian distribution with a mean of zero, this term will also have a mean of zero. 

Using the above properties, the \textit{reduced} validation $\chi^2$ becomes simply
\begin{equation}\label{simplified}
\frac{\chi^2_{\mathrm{val}}}{N_\tau} \coloneqq \frac{1}{N_\tau K}\sum_{k=1}^{K}\sum_{i=1}^{N_\tau} \left( \frac{G(\tau_i) - {G}_{\rm ex}(\tau_i)} {\sigma_k(\tau_i)}\right)^2 + 1.
\end{equation}

When preforming cross validation, it is important to consider the error level of the individual sampling and validation data sets. The number $K$ must be chosen such that the assumptions made in deriving Eq.~\eqref{simplified} are valid, but should be small enough to maintain a low enough error level. As displayed in the many examples investigated in Ref.~\onlinecite{shao_23}, an error level of $\sigma \sim 10^{-5}$ is typically sufficient to accurately reproduce the underlying spectra. When the spectral function contains many fine features, such as collections of small peaks or sharp edges, a smaller error level may be necessary to accurately resolve said features. However, for the spectra we consider here, this is not the case, so we have used data with $\sigma \sim 10^{-5}$ in all of the examples shown in the following sections. For the tests using QMC-generated data, this restricted us to using $K=20$ validation data sets. For consistency, we have also used this number of validation sets for our tests using synthetic data, where we have direct control of both the error level and number of $G(\tau)$ bins. 

In Appendix~\ref{half_sec}, we test an alternative cross validation procedure that maintains as high a data quality as possible in both the sampling and validation sets, splitting the $G(\tau)$ data into just two mutually exclusive sets.

\section{Optimal Sampling Temperature}\label{temp}
\subsection{Single Gaussian Peak}\label{temp_A}
Our first implementation of cross validation will be to provide support for the optimal $\Theta$ criterion, Eq.~\eqref{criterion}. The motivation behind this criterion is to achieve a statistically good data fit while placing the simulation in a temperature range safely above the regime of over-fitting, which we will be able to identify by tracking the behavior of the validation $\chi^2$ as $\Theta$ is lowered. 

For our tests, we will use synthetic $G(\tau)$ data aimed to mimic the data generated in a QMC simulation. Given an artificial spectrum $S(\omega)$, we use Eq.~\eqref{G_tau} to calculate the exact imaginary-time correlation function. To introduce the statistical errors necessarily present in QMC-generated data, we generate many $G(\tau)$ bins with noise that is correlated in imaginary-time. This is done by first generating normally distributed noise for each $\tau$ point $\varepsilon_0(\tau_i)$, and then taking the weighted averaging over all $\tau$ points with an exponentially decaying weight function:
\begin{equation}
\varepsilon(\tau_i) = \sum_{j=1}^{N_\tau} \varepsilon_0(\tau_j) e^{-|\tau_i - \tau_j|}.
\end{equation}
It was shown that the presence of covariance in the data actually improves the results of the SAC method \cite{shao_23}, so including this in our synthetic data is critical when comparing to QMC-generated data.

In our first test, we consider a spectrum consisting of a single Gaussian peak, centered at $\omega=3$. This spectrum can be essentially perfectly reproduced using the unconstrained SAC procedure, Fig.~\ref{params}(a) and (b). Here,  $S(\omega)$ is represented as a sum of $N_\omega$ $\delta$-functions that can freely move around in frequency-space,
\begin{equation}\label{free_samp}
S(\omega) = \sum_i A_i \delta(\omega - \omega_i),
\end{equation}
in some cases with $A_i$ variable and sampled  \cite{beach_04}, as depicted in Fig.~\ref{params}(b), but in this case we use fixed amplitudes $A_i = 1/N_\omega$, Fig.~\ref{params}(a). In this example, we use $N_\omega = 5,000$ total $\delta$-functions, which is large enough to no longer detect the dependence of $S(\omega)$ on $N_\omega$.

We generated $K = 20$ independent validation data sets, each containing 1,000 bins of synthetic $G(\tau)$ data with an error level $\sigma = 10^{-5}$. While it is certainly reasonable to generate this volume of data using modern QMC methods, this would be an inefficient use of computational resources given that the majority of the data is being used for validation. However, we used this large quantity so that the simplifications made in the large $K$ limit are more accurately realized. As we show in Sec. \ref{constrained}, one can still use cross validation with far fewer $G(\tau)$ bins, even when this limit is not fully reached.

When converting from $S(\omega)$ to $G(\tau)$, we set the inverse temperature $\beta = 2$ and use a $\tau$ spacing of $\Delta_\tau = 0.0625$, giving $N_\tau = 16$ imaginary time points. The relatively low value of $\beta$ used here was chosen in light of the fact that spectra consisting of broad and smooth features, such as the simple Gaussian used here, would usually be found at higher temperatures.

\begin{figure}[t]
    \centering
    \includegraphics[width = .8\columnwidth]{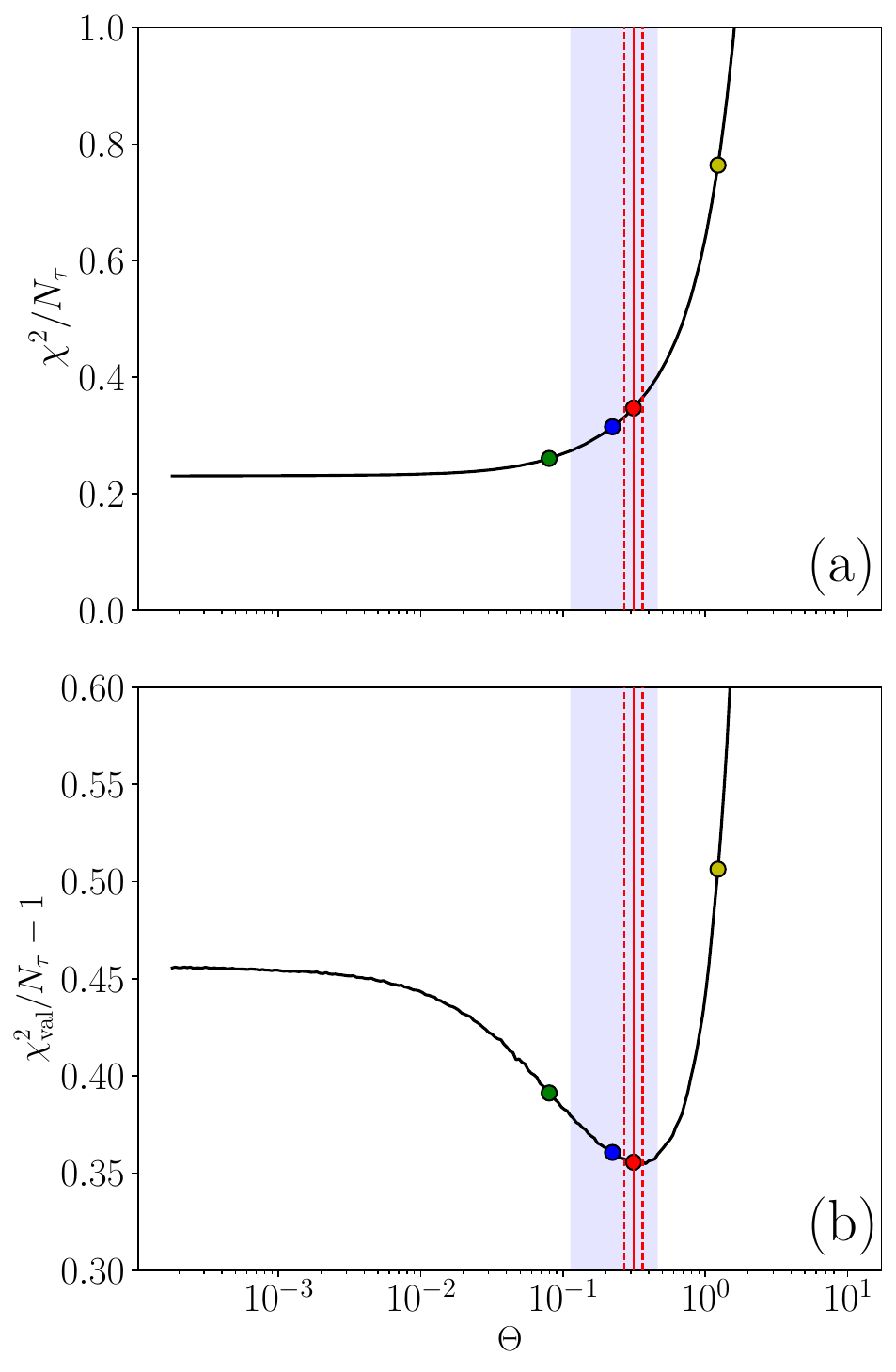}
    \caption{For the spectral function shown in Fig.~\ref{sig5_spec}, the average of $K=20$ cross validation annealing runs using synthetic data with a noise level of $\sigma = 10^{-5}$. Panels (a) and (b) show the behavior $\chi^2$ and  $\chi^2_{\mathrm{val}}$ versus $\Theta$. The temperature range corresponding to Eq.~\eqref{criterion} with $a=0.25-1.0$ is shaded in blue. The temperature where  $\chi^2_{\mathrm{val}}$ is at its minimum is marked by a red vertical line. The spread (one standard deviation) in the location of the $\chi^2_{\mathrm{val}}$ minimum is denoted by the red dashed lines. The $\chi^2$ and $\chi^2_{\mathrm{val}}$ values have been normalized by the number of $\tau$ points, and the background value of 1, corresponding to $X_2$ in Eq.~\eqref{chi2_val_factor} has been subtracted from $\chi^2_{\mathrm{val}}$. The colored points along the curves mark the sampling temperatures of the corresponding spectra in Fig.~\ref{sig5_spec}}
    \label{sig5_anneal}
\end{figure}

\begin{figure}[t]
    \centering
    \includegraphics[width = \columnwidth]{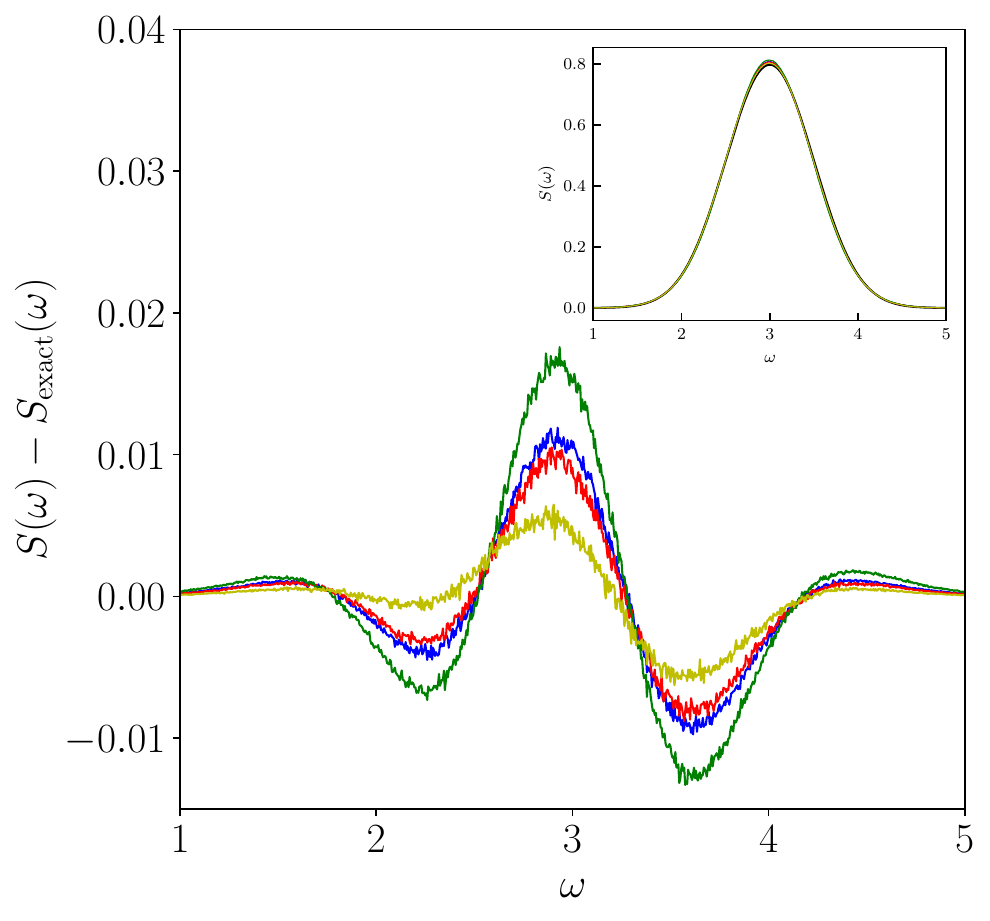}
    \caption{Deviation between the exact spectrum and those sampled at four different temperatures $\Theta$: that where $a=0.5$ (blue), where $\chi^2_{\mathrm{val}}$ is minimized (red), and $\Theta$ values above/below (yellow/green) this minimum. Inset: the spectral function themselves. In this case, it is difficult to distinguish the different spectra, as they nearly overlap.}
    \label{sig5_spec}
\end{figure}

The results of the cross validation procedure are presented in Fig.~\ref{sig5_anneal} and Fig.~\ref{sig5_spec}. Fig.~\ref{sig5_anneal} show the behavior of the sampling $\chi^2$ (panel (a)) and the validation $\chi^2$ (panel (b)) versus $\Theta$, averaged over all $K=20$ validation data sets and normalized by the number of $\tau$ points. The temperature range corresponding to Eq.~\eqref{criterion} with $a=0.25-1.0$ is shaded in blue. The temperature where  the validation $\chi^2$ is at its minimum is marked by a red vertical line, with the red dashed lines denoting the standard deviation of the location of the minimum. In practice, it is useful to repeat this process, rotating which of the $K+1$ data sets is used for sampling and which are used for validation, and averaging the results in the end. However, for this test we found this this was not necessary given the large number of synthetic data bins that can be easily generated. 

In Fig.~\ref{sig5_spec}, we show the difference between the the exact spectrum and the SAC spectra sampled at the temperature where $a=0.5$ (blue), at the validation $\chi^2$ minimum (red), and at temperatures above (yellow) and below (green) the validation minimum, for comparison. In the inset of Fig.~\ref{sig5_spec}, we show the spectral functions themselves. To generate these spectra we ran the SAC procedure using all bins, recombined from the $K+1 = 21$ sets, thus taking full advantage of the data at hand.

In this case, the minimum of the validation $\chi^2$ is quite sharp and agrees very well with the temperature from Eq.~\eqref{criterion}. Consequently, the red and blue spectra in Fig.~\ref{sig5_anneal} are nearly identical, and, for all intents and purposes, are equally valid representations of the true spectrum, shown in black. The green and yellow spectra are included to show the relatively small $\Theta$ dependence for this simple spectrum.

\begin{figure}[t]
    \centering
    \includegraphics[width = .8\columnwidth]{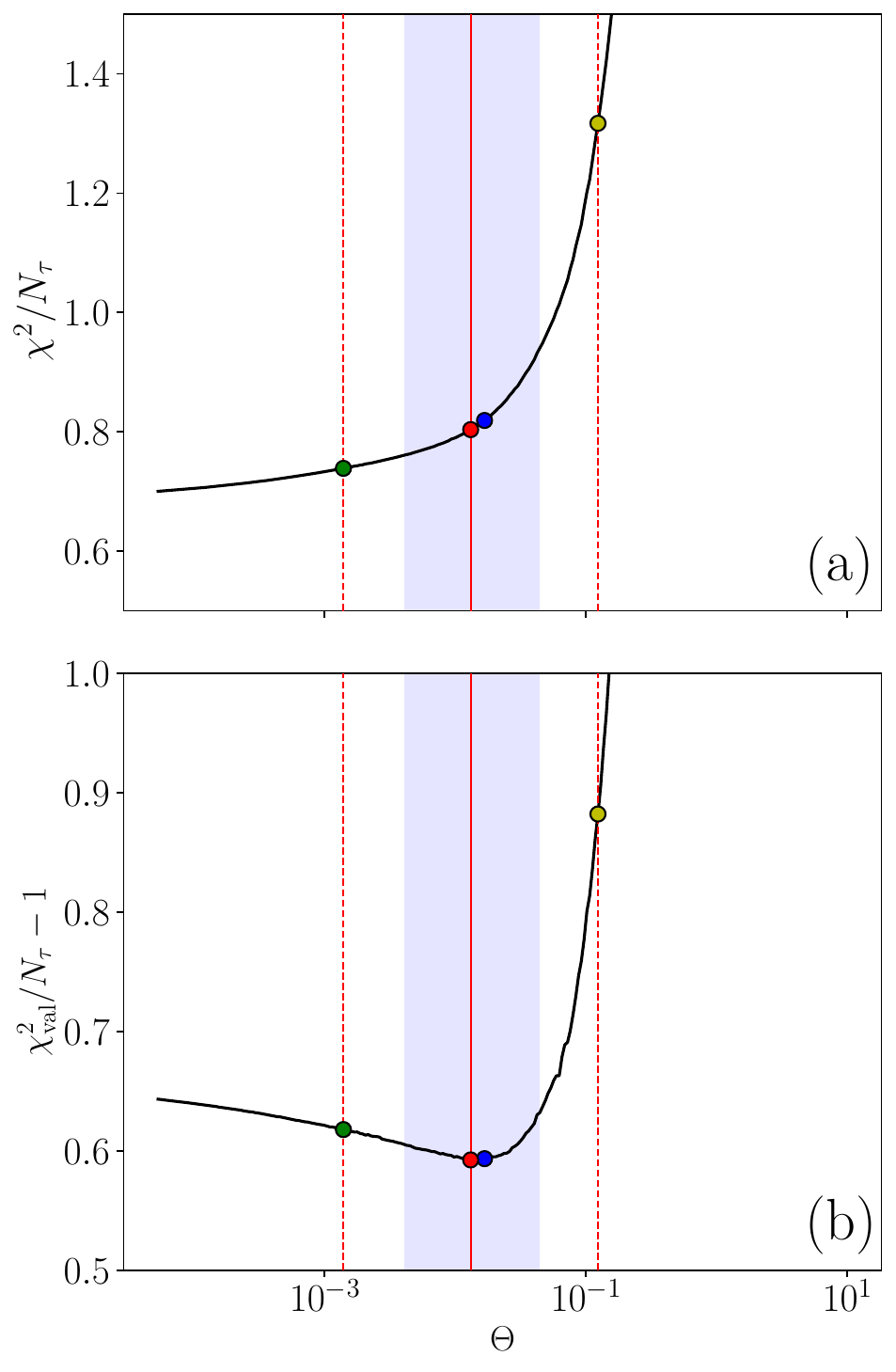}
     \caption{For the spectral function shown in Fig.~\ref{sig5_double_spec}, the average of $K=20$ cross validation annealing runs using synthetic data with a noise level of $\sigma = 10^{-5}$. Panels, symbols, and set up are identical to those in Fig.~\ref{sig5_anneal}.}
    \label{sig5_double_anneal}
\end{figure}
\subsection{Two Gaussian Peaks}\label{temp_B}
We now turn to an example where the unconstrained sampling parameterization in Eq.~\eqref{free_samp} cannot reproduce the artificial spectrum quite as closely. The spectrum we consider is composed of a pair of Gaussian peaks, one broader than the other, that intersect slightly. The area near the sharp dip, where the two peaks overlap, poses difficulties and would require a lower error level than one could reasonably achieve using real QMC-generated data to resolve fully \cite{shao_23}. As such, the criterion in Eq.~\eqref{criterion} provides an estimate for the temperature corresponding to the best possible spectrum, given these limitations.

\begin{figure}[t]
    \centering
    \includegraphics[width = \columnwidth]{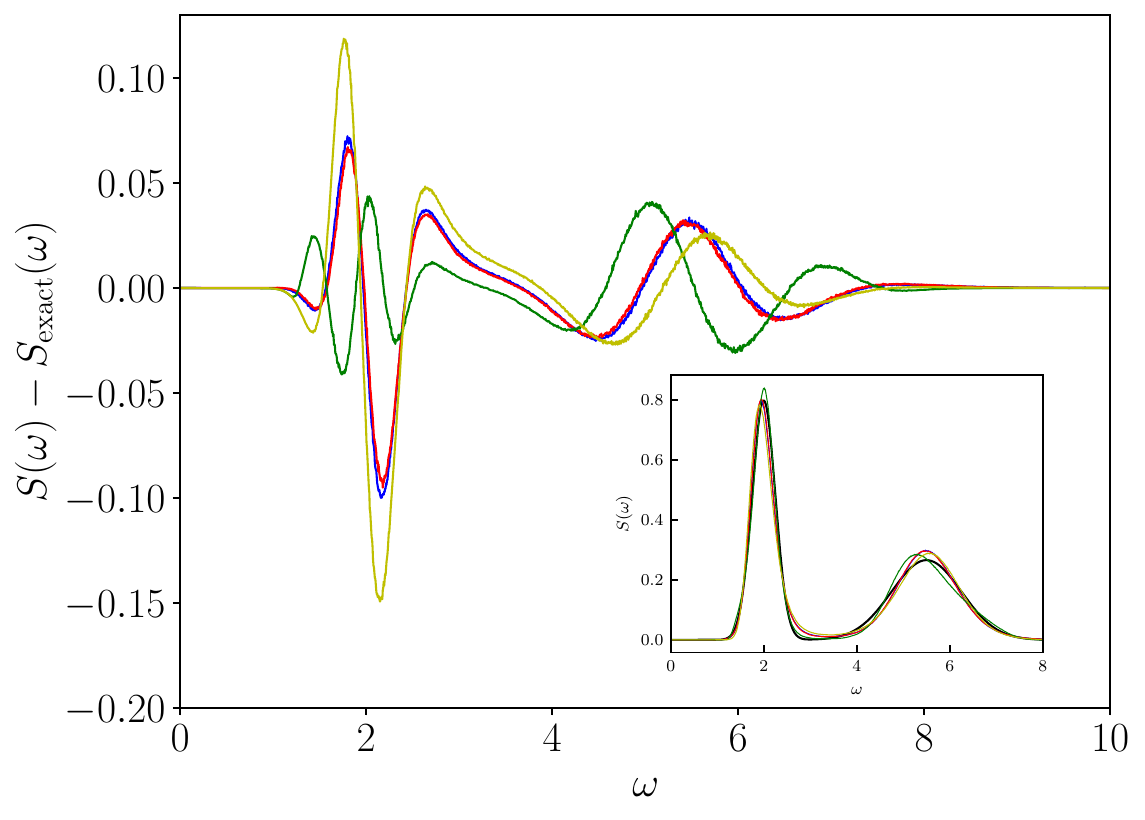}
    \caption{Deviation between the exact spectrum and those sampled at three different temperatures: the temperature where $a=0.5$ (blue), the temperature where $\chi^2_{\mathrm{val}}$ is minimized (red), at a slightly lower temperature, corresponding to one standard deviation below this minimum (green), and at a slightly higher temperature, corresponding to one standard deviation above the minimum (yellow). We note that the magnitude of the deviation is both a reflection of the error in the amplitudes and the locations  of the two Gaussian peaks. Inset: the spectral function themselves. Just as in Fig.~\ref{sig5_spec}, the red and blue spectra are nearly indistinguishable.}
    \label{sig5_double_spec}
\end{figure}

We performed the same test as in Sec. \ref{temp_A}, and present the results in Fig.~\ref{sig5_double_anneal}. In this case, both the validation minimum and the temperature range corresponding to the criterion with $a$ ranging from 0.25 and 1 are much broader, but still align quite well with one another. The higher level of uncertainty is a reflection of the inability of the unconstrained sampling scheme to reproduce this spectrum nearly as well as in the previous case, which can be seen clearly in Fig.~\ref{sig5_double_spec}. Here we again plot the spectra at the validation $\chi^2$ minimum (red) and at the temperature where $a=0.5$ (blue), as well as at a slightly lower/higher temperature, corresponding to one standard deviation below/above from the validation minimum (green/yellow). It is interesting to note that, at the lower sampling temperature (green), SAC is able to reproduce certain features of the exact spectrum better than at the temperatures corresponding to the validation minimum or $a=0.5$ (such as the position of the first peak and area where the two peaks intersect), while reproducing others not as well (such as the position of the second peak). This is consistent with the  broadness of the validation $\chi^2$ minimum, as it is not necessarily clear which is the better choice of spectra. Furthermore, all of the spectra shown here are acceptable reproductions of the underlying artificial spectrum; the most prominent features of the spectrum are captured well in all four cases and the deviations plotted in Fig.~\ref{sig5_double_spec} reflect only small shifts in the locations and amplitudes of the two Gaussian peaks. 

We emphasize that using cross validation to determine the optimal sampling temperature is not necessary when running SAC with real QMC-generated data. Rather, these test-cases show that the criterion in Eq.~\eqref{criterion} indeed samples safely above the regime of over-fitting, while still keeping $\chi^2$ at an acceptable level, and \textit{should} be used as the method of fixing $\Theta$ in real use-cases. In the following sections we will explore a more practical and decisive use for our cross validation procedure, as a tool for selecting the most applicable SAC parameterization.


\section{Optimal Spectral Parameterization: Unconstrained Sampling}
Our second application of cross validation will be to help determine which SAC parameterization to use when the exact spectral features are unknown, i.e. model selection \cite{bishop_06, mehta_19}. In Eq.~\eqref{free_samp}, we model the spectrum as unconstrained $\delta$-functions with sampled positions (Fig.~\ref{params}(a)). In addition to the positions of the $\delta$-functions, their amplitudes may also be variable and sampled, as depicted in Fig.~\ref{params}(b). These two parameterizations, which we'll refer to as equal and variable amplitudes (EA and VA, respectively), have different configurational entropies as well as different sampling efficiencies, depending on the exact form of the underlying spectrum. For the EA parameterization, the entropy is the conventional Shannon information entropy \cite{beach_04, shao_23}, while for the VA parameterization, a generalized R\'enyi entropy is instead appropriate \cite{ghanem_23}.

It was shown that if these entropies are used in MEM, the results will be identical to those from SAC, when using the corresponding unconstrained parameterization, with the caveats that $N_\omega$ must be large and the MEM entropy regulator $\alpha$ must be chosen such that the two output spectra have the same $\chi^2$ values (the choice of $\alpha$ when using the variable amplitude parameterization is slightly more subtle due to the specifics of the entropy form used) \cite{shao_23}.

In many cases, such as the two considered in the previous section, the output spectra will depend very little, if at all, on whether amplitudes are sampled along with frequencies. More specifically, if $N_\omega$ is large enough, the spectral features are not very sharp, and the error level is low, the different entropy contents will only minimally affect the sampling efficiency and the underlying spectrum can be reproduced with high fidelity using either parameterization \cite{shao_23}. But this is not always the case, as we will explore in the following sections.

\subsection{Two Equal Amplitude Gaussian Peaks}\label{double_peak}

\begin{figure}[t]
\begin{center}
\includegraphics[width=\columnwidth]{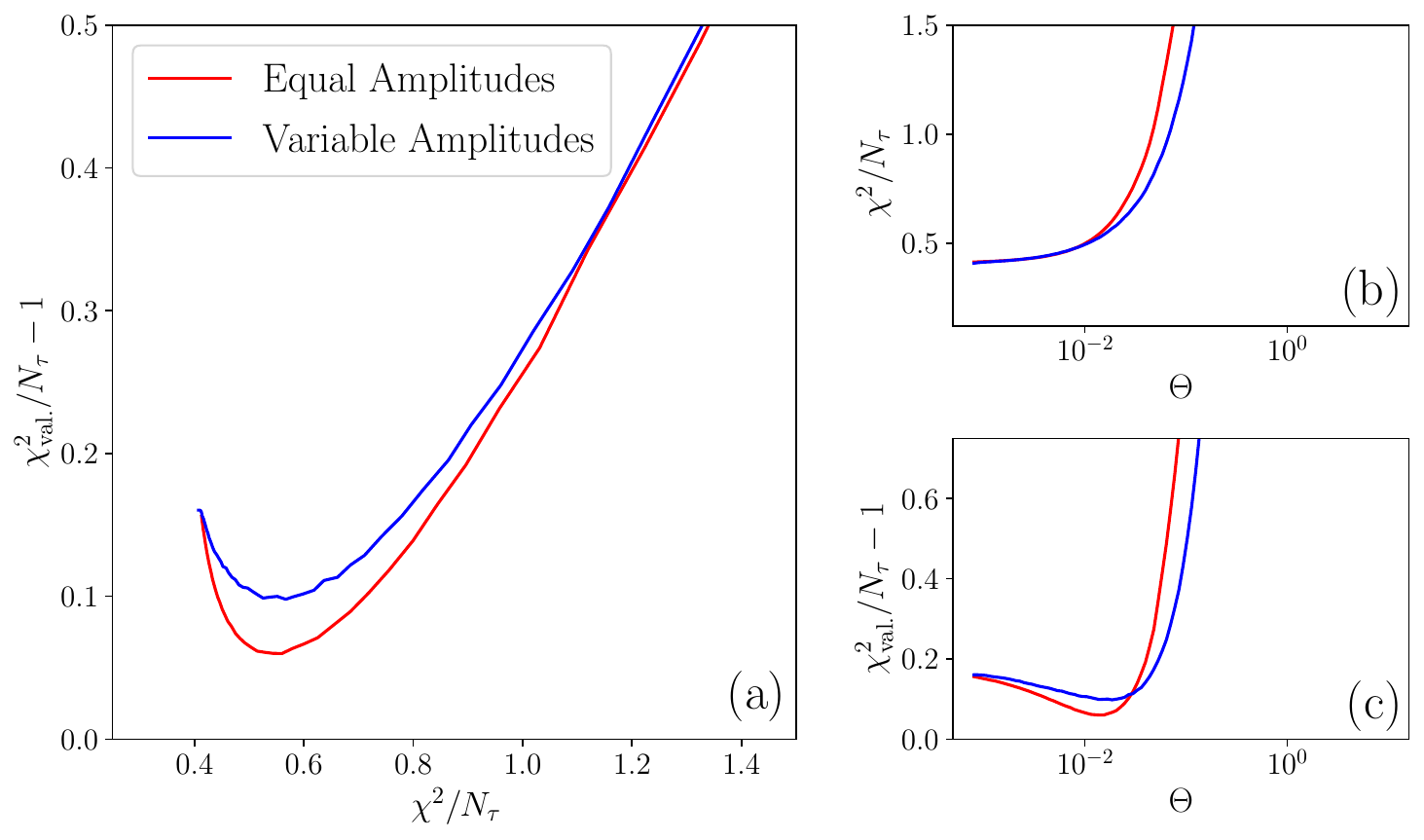}
\caption{Cross validation results for the spectral function shown in Fig.~\ref{double_peak_spec}. In panel (b), the sampling $\chi^2$ values, normalized by the number of $\tau$ points, is plotted versus the sampling temperature $\Theta$. In panel (c), the validation $\chi^2$ values, also normalized by the number of $\tau$ points, but now also averaged over all cross validations data sets, is plotted versus the sampling temperature $\Theta$. We have subtracted the background constant of one from the validation $\chi^2$ to account for the term $X_2$ in Eq.~\eqref{chi2_val_factor}. In panel (a), the validation $\chi^2$ as a function of the sampling $\chi^2$ is plotted to allow for direct comparison of the two parameterizations at a fixed value of the sampling $\chi^2$.}
\label{double_peak_anneal}
\end{center}
\end{figure}

\begin{figure}[t]
\begin{center}
\includegraphics[width=\columnwidth]{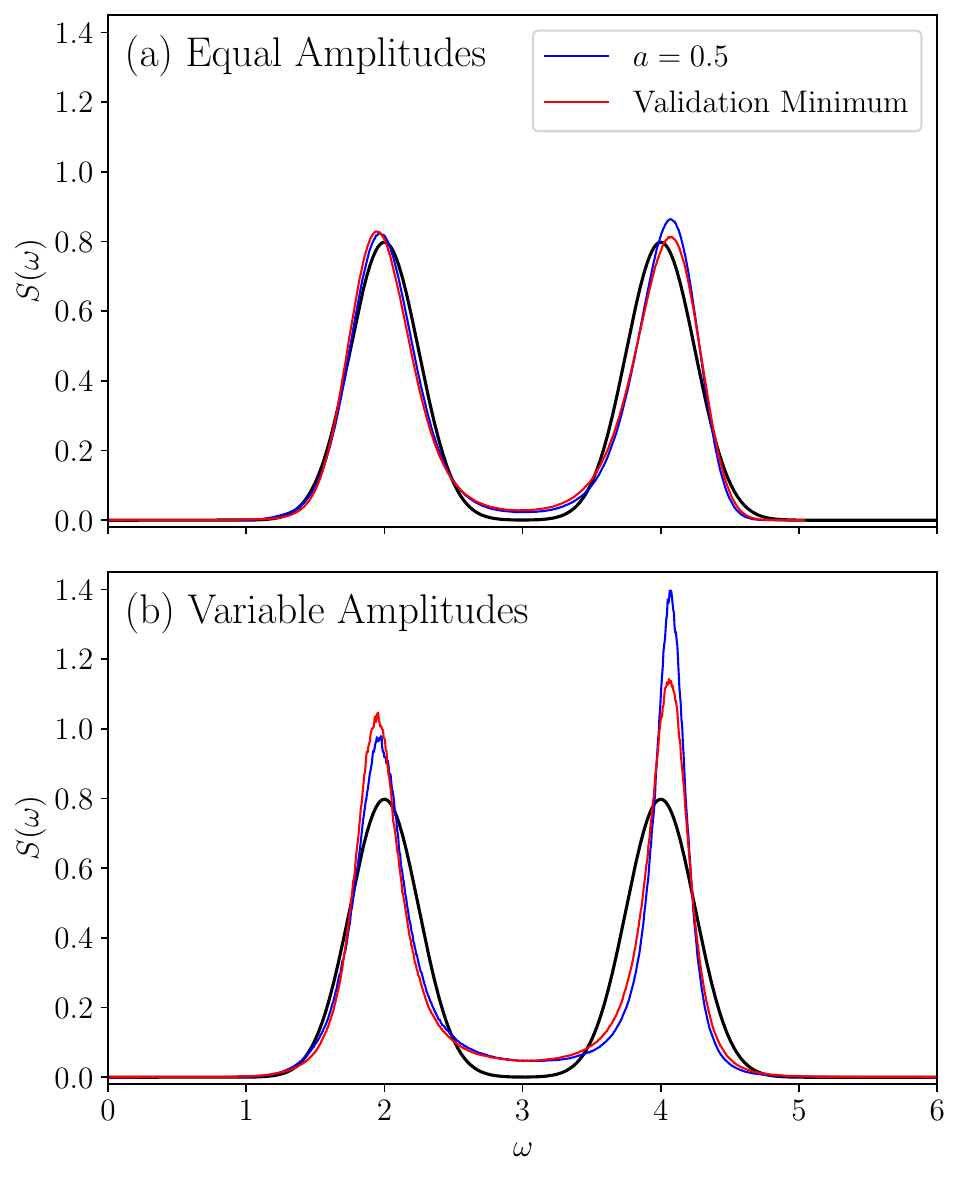}
\caption{Spectra corresponding to the two parameterizations shown in Fig.~\ref{double_peak_anneal}, equal amplitudes (panel (a)) and variable amplitudes (panel (b)). The exact spectrum is plotted in black, while the two colored spectra correspond to two different sampling temperatures: the validation $
\chi^2$ minimum (red) and the criterion in Eq.~\eqref{criterion} with $a=0.5$ (blue).  In both cases, $N_\omega$ = 4,000 $\delta$-functions were used, which was chosen to ensure that there was no dependance of $S(\omega)$ on $N_\omega$.}
\label{double_peak_spec}
\end{center}
\end{figure}

In the first case we consider, we used an artificial spectrum composed of two equal amplitude Gaussian peaks centered at $\omega = 2$ and $\omega = 4$. While this spectrum is relatively simple, the proximity of the two peaks poses similar difficulties to those encountered in Sec.~\ref{temp_B}; the low amplitude region in between the two peaks cannot be resolved fully at this error level, which causes a ripple effect of distortions throughout the spectrum. The two unconstrained sampling schemes, EA and VA, are able to overcome these challenges to different extents, which can be seen clearly when performing tests using artificial spectra. However, when performing analytic continuation using QMC-generated data, one must instead rely on cross validation to asses the abilities of these two parameterizations. 

Just as in Sec.~\ref{temp}, we generated $K+1 = 21$ sets of synthetic data with error level $\sigma = 10^{-5}$. In this case, we used a slightly higher inverse temperature, $\beta = 16$, and accordingly a larger $\tau$ spacing, $\Delta \tau = 0.1$, when converting the artificial spectrum to $G(\tau)$. When defining the error level, we introduce a cutoff of $\tau_{\rm max} \approx 4$, chosen so that the relative error on $\bar{G}(\tau)$ does not exceed 10\%. We ran the cross validation procedure using two different SAC parameterizations, the unconstrained sampling method with equal amplitude and variable, sampled amplitude $\delta$-functions, and compare their performance.

Figure~\ref{double_peak_anneal} shows the results of this cross validation test. The normalized sampling $\chi^2$ during the annealing run is shown in panel (b) for both updating methods, equal (red) and variable (blue) amplitudes. Because the entropy contents of the spectra defined with each parameterization are different, the annealing paths they take differ, so it is also informative to compare the validation $\chi^2$ values at the same value of the sampling $\chi^2$. This is shown in Fig.~\ref{double_peak_anneal}(a) where we plot the validation $\chi^2$ versus the sampling $\chi^2$ throughout the annealing run.

For this spectrum, the validation $\chi^2$ reaches the lowest minimum value for the EA parameterization, which can be clearly seen in Fig.~\ref{double_peak_anneal} panel (c), and even more clearly in panel (a). We can see why this is the case by comparing the spectra produced using both parameterizations, shown in Fig.~\ref{double_peak_spec}. Here, we plot the spectra at two sampling temperatures, corresponding to the validation $\chi^2$ minima (red, with $\chi/N_{\tau} \sim 0.56$) and the criterion in Eq.~\ref{criterion} with $a = 0.5$ (blue, with $\chi/N_{\tau} \sim 0.54$).
 

The EA parameterization (panel (a)) produces a spectrum where both the shapes and heights of the two Gaussian peaks are very close to those in the exact spectrum, shown in black. On the other hand, the VA parameterization (panel (b)) produces peaks that are far too sharp with amplitudes that are unequal and too large. While both parameterizations are unable to correctly reproduce the region in between the two peaks, the VA parameterization performs far worse. The excessively tall peaks produced by the variable amplitudes parameterization may be compensation for the missing weight in this region.

Each SAC parameterization comes with its own set of entropic pressures associated with the degrees of freedom that are being sampled. In the case of the VA parameterization, it is known that these pressures overly favor sharp peaks \cite{shao_23}. In this example, that comes as a detriment, but in others, this can be advantageous, as we will explore in the following section.

\subsection{Sharp Gaussian Peak with a Shoulder}\label{shoulder}

\begin{figure}[t]
\begin{center}
\includegraphics[width=\columnwidth]{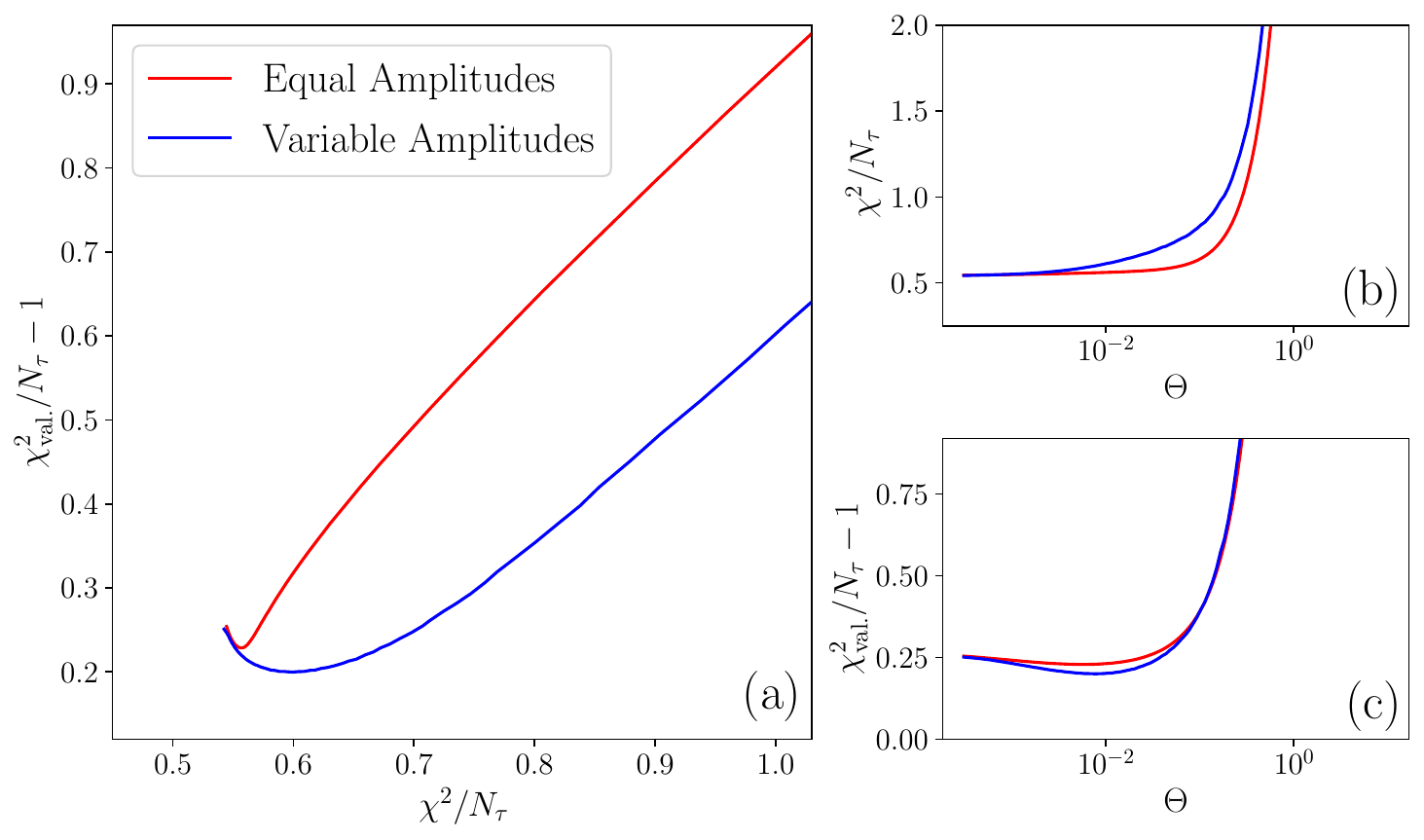}
\caption{Cross validation results for the spectral function shown in Fig.~\ref{shoulder_spec}. Panels and set up are identical to those in Fig.~\ref{double_peak_anneal}}
\label{shoulder_anneal}
\end{center}
\end{figure}

For our second test, we consider an artificial spectrum composed of a tall and sharp Gaussian peak centered at $\omega = 2$, followed by a short and broad Gaussian at $\omega=3$, which forms a shoulder-like feature for the first peak. A similar artificial spectrum was used in Ref.~\onlinecite{shao_23} to study this exact issue, whether or not to sample amplitudes along with frequencies. It was found that including amplitude updates was necessary to resolve both the dominant peak and the shoulder for data with error levels attainable using QMC \cite{shao_23}. Being able to determine this using cross validation would be very valuable when the exact underlying spectrum is unknown.

\begin{figure}[t]
\begin{center}
\includegraphics[width=\columnwidth]{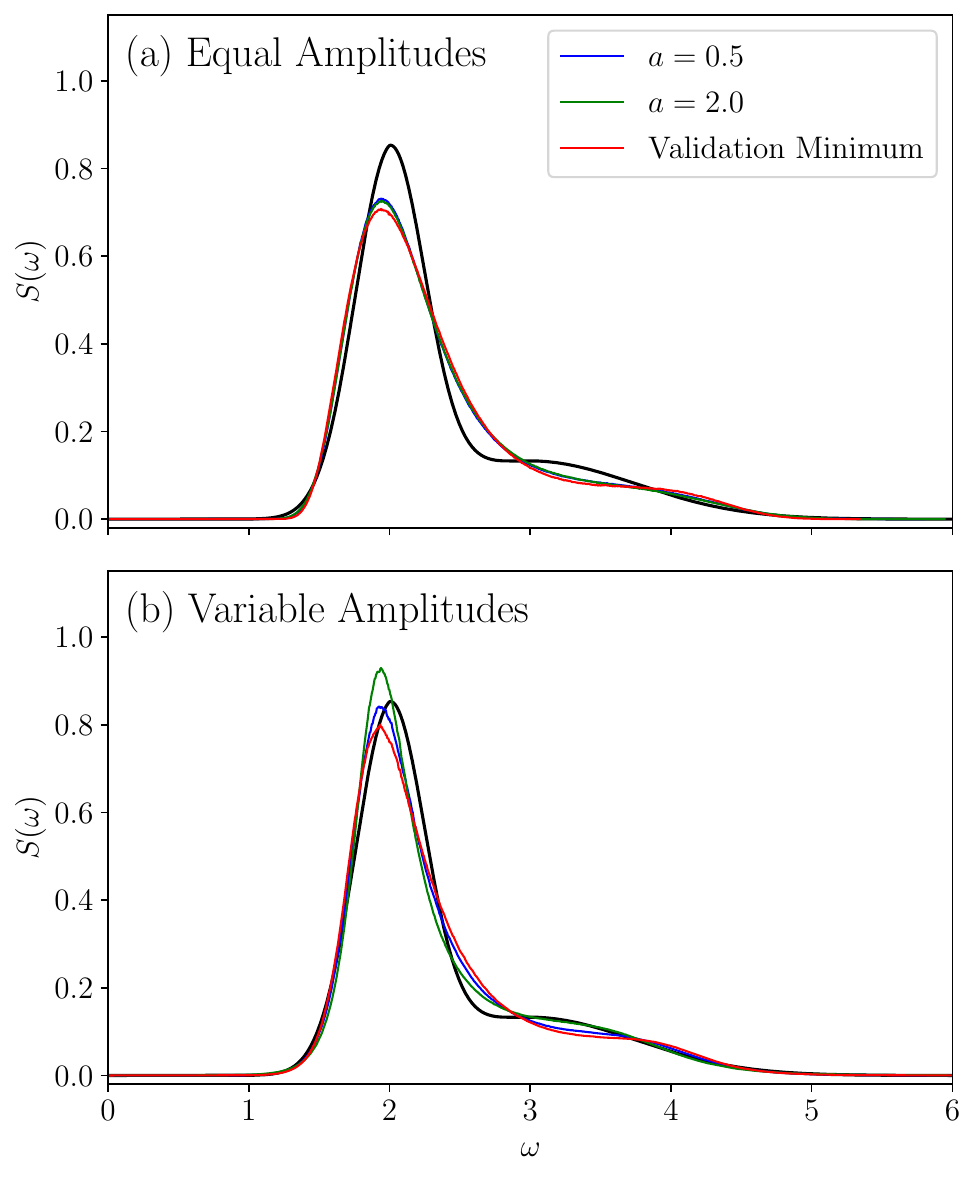}
\caption{Spectra corresponding to the two parameterizations shown in Fig.~\ref{double_peak_spec}. Panels and set up are identical to those in Fig.~\ref{double_peak_spec}, except an additional spectrum is plotted, corresponding to the criterion in Eq.~\eqref{criterion} with $a=2.0$.}
\label{shoulder_spec}
\end{center}
\end{figure}

Again, we generated $K+1 = 21$ sets of synthetic data with error level $\sigma = 10^{-5}$ and use $\beta = 16$ with $\Delta \tau = 0.1$ (again, with $\tau_{\rm max} \approx 4$), which is the same inverse temperature used in Ref.~\onlinecite{shao_23} for their tests on this type of spectral function. The results of the cross validation procedure are shown in Fig.~\ref{shoulder_anneal} and the corresponding spectral functions are shown in Fig.~\ref{shoulder_spec}.

In this case, cross validation suggests that the VA parameterization performs better, in contrast to the previous example. Here, the validation $\chi^2$ reaches a lower minimum value (panel (c)), but is also lower for the majority of the annealing process, when compared at fixed values of the sampling $\chi^2$ (panel (a)). 

In Fig.~\ref{shoulder_spec} we plot the spectra at three different sampling temperatures for both parameterizations. The red spectra were sampled at the temperature corresponding to the validation minima, and the blue and green spectra were sampled at temperatures corresponding criterion in Eq.~\eqref{criterion} with $a=0.5$ and $a=2.0$, respectively. 

At all three sampling temperatures, the VA parameterization is able to resolve the first peak with higher fidelity. In both cases, the spectra sampled at the validation $\chi^2$ minima (red, with $\chi/N_{\tau} \sim 0.55$) do not reproduce the shoulder feature very well, instead displaying a ringing pattern around this structure. But as the temperature, and consequently the $\chi^2$ value, are increased (blue, with $\chi/N_{\tau} \sim 0.66$ and green, with $\chi/N_{\tau} \sim 1.0$), the shoulder is reproduced with much higher accuracy by the VA parameterization. This is, however, not the case for the EA parameterization, as suggested by the results shown in Fig.~\ref{shoulder_anneal}(c). 

In the previous examples, there was much better agreement between the validation $\chi^2$ minima and the criterion in Eq.~\eqref{criterion} than there is for this case. We explain this discrepancy by noting that at the error levels considered here, $\sigma = 10^{-5}$, this spectrum is unable to be produced with nearly as high a fidelity. If the SAC results will always deviate from the exact solution in some considerable way, then the validation $\chi^2$ may be minimized by a spectrum that does not visually appear to match the exact spectrum quite as well as one that is produced at a slightly higher or lower sampling temperature. This is exemplified by the $a=2.0$ spectrum in Fig.~\ref{shoulder_spec}(b); while it may visually appear to be the most accurate, the results of the cross validation procedure suggests that this is not the case. Cross validation identifies the spectrum with the corresponding $G(\tau)$ that is closest to the exact solution, and in some cases this may be a spectrum that deviates from the exact $S(\omega)$ in multiple different ways, i.e. a compensating effect that produces multiple different distortions. Even considering these limitations, it is clear that cross validation can accurately determine which unconstrained sampling scheme produces the most accurate spectrum.

At this point it should be pointed out that in the four unconstrained sampling tests we have presented, the fidelity of the SAC spectrum varies depending on the exact form of the underlying spectrum. In light of this, the cross validation $\chi^2$ value should not be taken as an absolute measure, but rather a relative statistic used to compare two (potentially imperfect) solutions. The ability of cross validation to identify which imperfect solution is more accurate, as gauged by direct comparison to the exact spectrum, is a testament to the robustness of this procedure.

\section{Optimal Spectral Parameterization: Constrained Sampling}\label{constrained}

There is second class of SAC parameterizations that have an even greater impact on the shape and form of the output spectra, which refer to as constrained SAC sampling schemes (Fig.~\ref{params}(c)-(e)). If certain restrictions are imposed on the locations and amplitudes of the $\delta$-functions, the SAC method can reproduce sharp spectral features present only at low temperatures, such as narrow quasi-particle peaks and power-law edge singularities \cite{sandvik_16, sandvik_17, shao_23}. In many cases, it is known that the spectral function contains such a sharp feature, but the exact form of this feature, and thus which constrained sampling scheme to use, is unknown. In the following sections we will show how cross validation can be used to identify the most appropriate constrained SAC parameterization.


A spectrum containing a dominant $\delta$-peak followed by a smooth continuum can be reproduced using the parameterization depicted in Fig.~\ref{params}(b) \cite{sandvik_17}
\begin{equation}\label{S_peak}
    S_{\rm peak}(\omega) = A_0\delta(\omega - \omega_0) + \sum_i A_i \delta(\omega - \omega_i),
\end{equation}
where $A_0 + \sum_i A_i = 1$ and the constraint $\omega_i > \omega_0$ for all $i$ is imposed during the sampling process. With this parameterization, the location of the peak, $\omega_0$, is sampled within the program, along with all of the other $\delta$-functions, but the weight of the macroscopic $\delta$-peak, $A_0$,  is fixed and optimized using a simple scan to find a $\chi^2$ minimum, as detailed in Appendix~\ref{A0_scan}. This type of spectral function is appropriate for describing the spectrum of the $S = 1/2$  AFM Heisenberg model in two-dimensions (2D) \cite{wallin_93}, where linear spin-wave theory predicts the presence of a macroscopic, ``single-magnon'' peak \cite{igarashi_92}, while an incoherent continuum at higher energy can be attributed to either multimagnon excitations or partially deconfined spinon excitations (both of which cannot be captured by convention spin-wave theory \cite{sandvik_17}). The SAC parameterization in Eq.~\eqref{S_peak} has been used to produce results for the spectral function of the aforementioned square-lattice Heisenberg model that agreed very well with results from inelastic neutron-scattering experiments on a material considered the best physical realization of this model \cite{sandvik_17}.

Another example of a sharp spectral feature that can only be resolved using a constrained SAC sampling parameterization is an edge singularity. In quantum many-body systems with fractionalized excitations, spectral weight may be spread over a range of energies, often with a power-law distribution that diverges at some frequency $\omega_q$, where $q$ is the total momentum of the fractionalized quasiparticle \cite{shao_23}. One of the most well know examples of this is the spectral function of the operator $O = S_q^z$ in the $S = 1/2$ AFM Heisenberg chain. Using the Bethe ansatz (BA) solution, it was shown that in the thermodynamic limit, this spectral function diverges as $(\omega-\omega_q)^{-1/2}$, with a logarithmic correction (except for at $q=\pi$, where the divergence is faster) \cite{maillet_05, caux_2005, affleck_06}.

A spectrum with this type of feature can be realized in SAC by restricting the locations of the sampled $\delta$-functions such that the distance between adjacent $\delta$-function's monotonically increases (Fig.~\ref{params}(c) and (d)) \cite{sandvik_16, shao_23}:
\begin{equation}\label{S_edge}
     S_{\rm edge}(\omega) = \sum_i A_i \delta(\omega - \omega_i),
\end{equation}
with the constraint $\omega_{i+1} - \omega_i > \omega_{i}-\omega_{i-1}$, as depicted in Fig.~\ref{params}(c). Due to entropic pressure, this parameterization naturally produces a spectrum where the mean amplitude density of the $\delta$-functions forms precisely an edge singularity with the asymptotic behavior $(\omega-\omega_q)^{-1/2}$. The location of the edge, $\omega_q$,  is not a fixed, user-inputted parameter that must be scanned over, like in the case of the leading $\delta$ weight $A_0$ in Eq.~\eqref{S_peak}. Rather, the location of the edge adapts naturally to the data during the sampling process. Away from this edge the data dictates the exact shape of the spectrum, smoothly transitioning from the asymptotic power-law form. 

It is also possible to resolve an edge diverging with any power $p$, $(\omega-\omega_q)^{p}$ with $p<0$,  by properly adjusting the amplitudes $A_i$ \cite{shao_23}. For an edge that diverge faster/slower than $S(\omega \to \omega_q) \propto (\omega-\omega_q)^{-1/2}$, i.e. $p$ less/greater than $-1/2$, the amplitudes of the $\delta$-functions increase/decrease as they approach the edge. This is depicted in Fig~\ref{params}(d) for an edge that decays with an exponent $p = -3/4$.


In most cases, one tests both the unconstrained and constrained parameterizations when performing analytic continuation on QMC-generated data. However, relying on the value of $\chi^2$ alone to compare the resulting spectra is not always reliable, since the SAC method can produce spectra with acceptable $\chi^2$ values even when the improper parameterization is being used (a consequence of the ``ill-posed'' problem). Our cross validation procedure can be used as an unbiased method not only to determine whether an unconstrained or constrained sampling scheme should be used, but also to determine which of the constrained sampling scheme produces the most statistically likely spectrum.

\begin{figure}[h]
    \centering
    \includegraphics[width = .9\columnwidth]{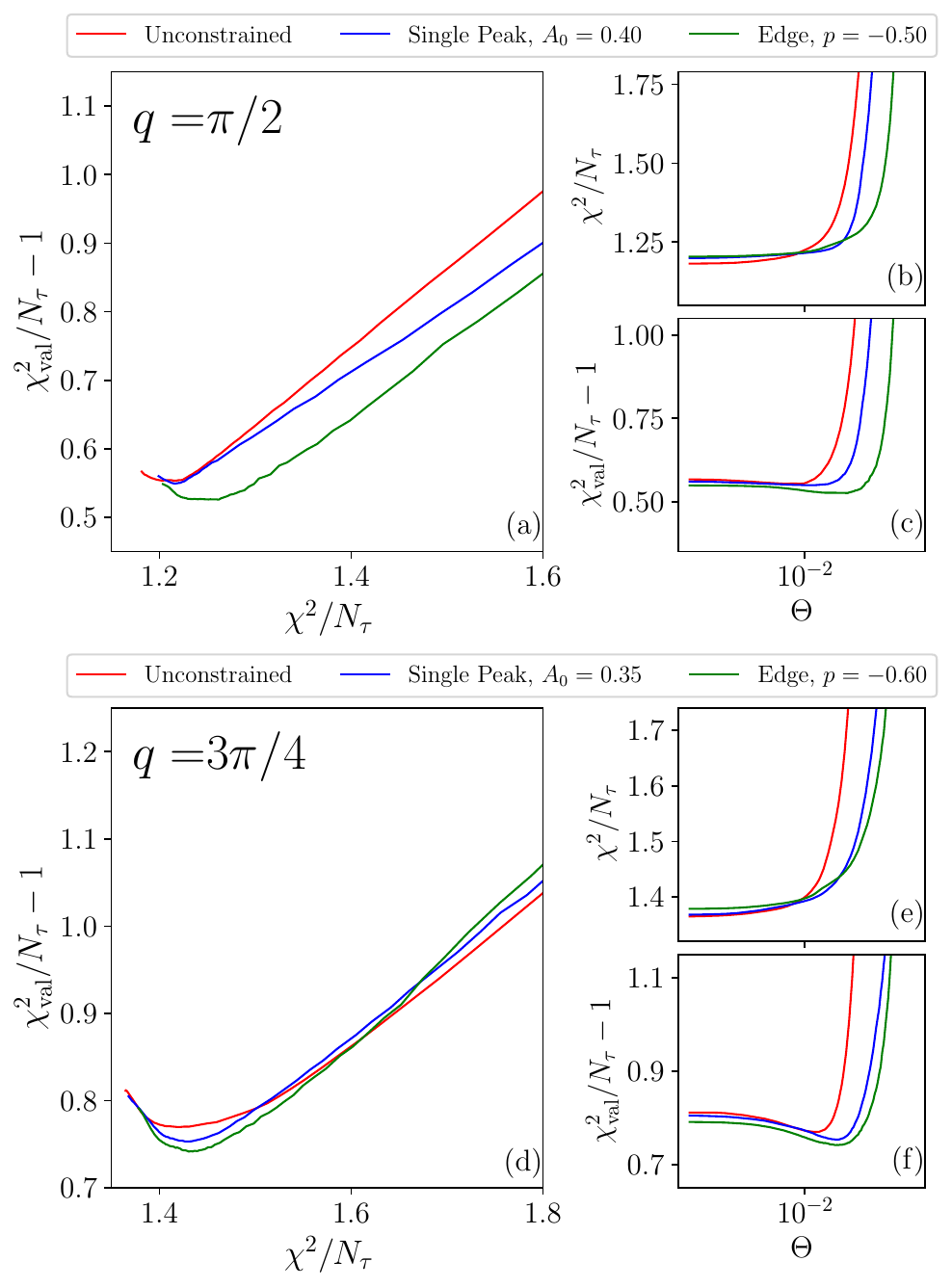}
    \caption{Cross validation results for $S(q=\pi/2, \omega)$ and $S(q=3\pi/4, \omega)$ of the $S=1/2$ Heisenberg chain ($L=512$, $\beta = 1024$). In panels (b) and (e), the sampling $\chi^2$ values, normalized by the number of $\tau$ points and averaged over all cross validations runs, is plotted versus the sampling temperature $\Theta$. In panels (c) and (f), the validation $\chi^2$ values, also normalized by the number of $\tau$ points (50 in this case) and averaged over all cross validations runs, is plotted versus the sampling temperature $\Theta$. We have subtracted the background constant of one from the validation $\chi^2$ to account for the term $X_2$ in Eq.\eqref{chi2_val_factor}. In panels (a) and (d), we plot the validation $\chi^2$ as a function of the sampling $\chi^2$, to allow for direct comparison of the three parameterizations at a fixed $\chi^2$ value.}
    \label{hchain_anneal}
\end{figure}

\subsection{Heisenberg Chain}\label{spec_param}

We first demonstrate the ability to discriminate between models using real QMC data, generated using the stochastic series expansion (SSE) method \cite{sandvik_10}, for the $S = 1/2$ AFM Heisenberg chain. We followed the procedure outlined in Sec. \ref{cross_val_proc}, again with $K=20$ validation sets with an error level of $\sigma = 10^{-5}$. The system size used here is $L=512$, and the QMC simulation was run using the inverse temperature $\beta=1024$, which is large enough to access ground state properties. We considered $S_q^z(\omega)$ at two momenta, $q = \pi/2$ and $3\pi/4$, as test cases. For both $q$ values, the optimal weights of the macroscopic $\delta$-peak were determined using the method detailed in Ref.~\onlinecite{shao_23} and shown here in Appendix~\ref{appendix_scans} ($A_0 =0.40$ and $0.35$ for $q = \pi/2$ and $3\pi/4$, respectively). Due to the logarithmic correction to the square-root divergent edge form of the spectral function for $O = S_q^z$  \cite{maillet_05, caux_2005, affleck_06}, the optimal power $p$ for each $q$ value considered may not necessarily be equal to $-0.50$. As we show in Appendix~\ref{p_scan}, for $q = 3\pi/4$, the optimal exponent is $p = -0.60$.

\begin{figure}[t]
    \centering
    \includegraphics[width = \columnwidth]{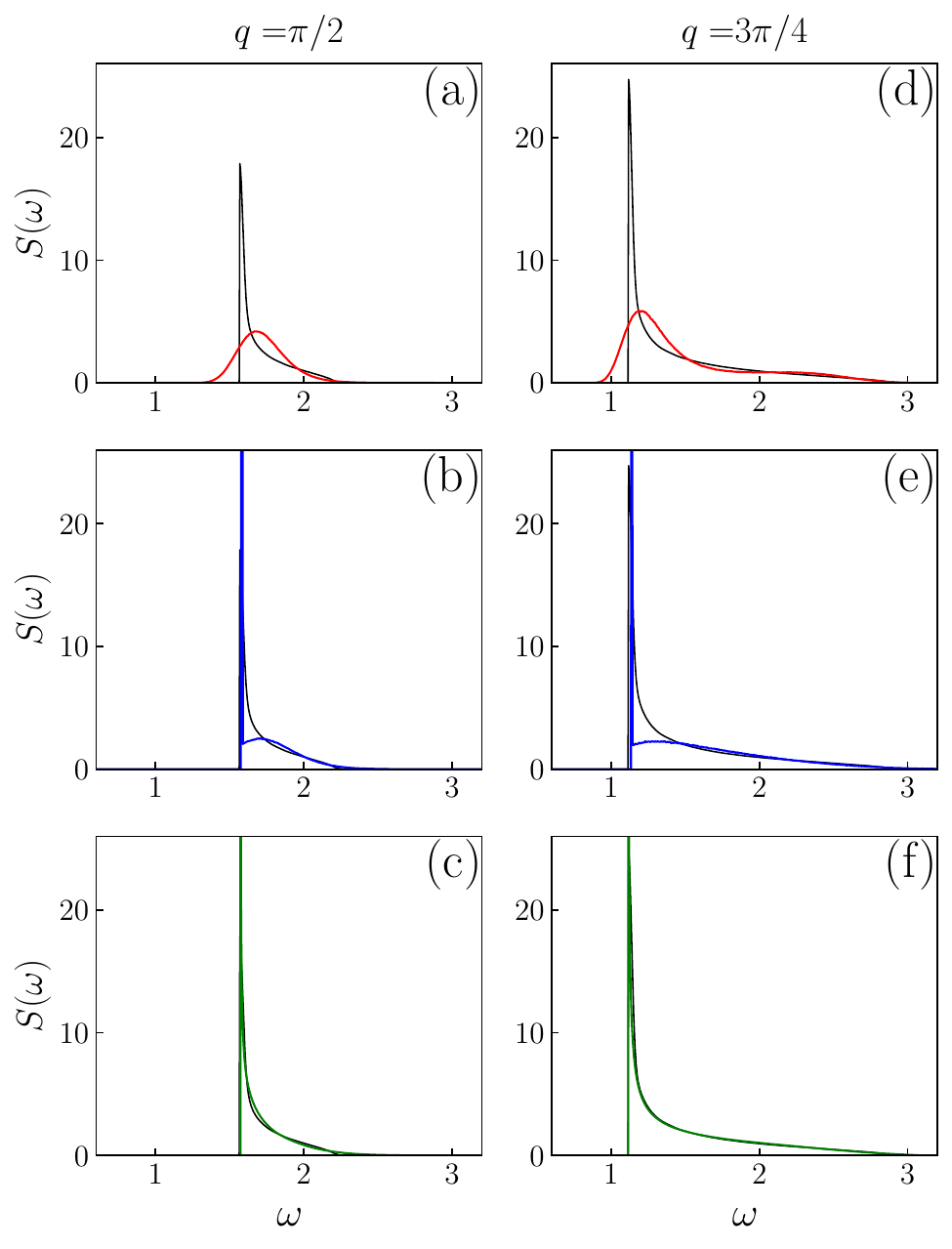}
    \caption{Spectra corresponding to the three parameterizations in Fig.~\ref{hchain_anneal}, unconstrained (a, d), single-peak with  (b, e), and edge (c, f). For each parameterization, the spectra are sampled at the temperatures corresponding to the criterion in Eq.~\eqref{criterion}, using their respective sampling $\chi^2$ curves. For the unconstrained sampling parameterization, we use $N_\omega = 2,000$ $\delta$-functions. For the single-peak parameterization we use the leading $\delta$-function weights $A_0$ shown in Fig.~\ref{hchain_anneal} and $N_c = 2,000$ continuum $\delta$-functions. For the edge parameterization, we use $N_\omega = 640$ and the exponent $p$ shown in Fig.~\ref{hchain_anneal}. In each panel, we also plot the BA results for a chain of length 500 (black curves), which have also been subject to smoothing \cite{caux_BA}, resulting in more rounded peaks.}
    \label{hchain_spec}
\end{figure}

For this test, we repeated the cross validation process for  three parameterizations (unconstrained, single-peak, and edge). We now also rotate which of the $K+1 = 21$ sets of QMC-generated data was used for sampling, and which ones were used for validation, and averaged the results over all of the rotations. We found that this extra step was necessary given the quality and quantity of the QMC-generated data at hand. 

Fig.~\ref{hchain_anneal} shows the results of this test. The the normalized sampling (panels (b) and (e)) and validation (panels (c) and (f)) $\chi^2$ values are plotted versus $\Theta$ for each parameterization: unconstrained (red), single-peak (blue), and edge (green). Just as with the two unconstrained sampling parameterizations, the entropy contents of the spectra defined with each constrained sampling parameterization are different, so to better compare the three parameterizations, we plot the validation $\chi^2$ versus the sampling $\chi^2$ as well, shown in panels (a) and (d).

While it is difficult to glean any information about the relative performance of the three parameterizations from the sampling $\chi^2$, the validation $\chi^2$ paints a much different picture. At each temperature, the the edge parameterization has the lowest validation $\chi^2$, followed by the single-peak case, and then unconstrained sampling. This would suggest that the edge is indeed the most appropriate parameterization, in agreement with the BA solution for this model. This conclusion can also be made when comparing the values of the validation $\chi^2$ at fixed $\chi^2$. For $q= \pi/2$, the edge clearly performs the best of the three parameterizations during the entire annealing run. For $q= 3\pi/4$, the validation $\chi^2$ for the edge parametrization only dips below that of the other two at the end of the annealing run, but clearly reaches the lowest minimum value. We thus find that it is important take into consideration both the minimum value of each validation $\chi^2$ curve and the path the curves take when comparing the various parameterizations.

We now compare the corresponding spectra, shown in Fig.~\ref{hchain_spec}, to see if the hierarchy suggested by the validation $\chi^2$ is borne out in the results of the analytic continuation process. As reflected by the relative differences between the validation $\chi^2$ curves, the single-peak and edge spectra share similar features, such as lower bounds near $\omega = 1.5$ ($q = \pi/2$) and $\omega = 1.0$ ($q = 3\pi/4$), while the unconstrained sampling produces spectra with considerable spectral weight below the edge. The BA results \cite{caux_BA}, shown in black, agree very well with the edge spectra (with edges that nearly perfectly align), moderately well for the peak spectra (but with some large deviations), and very poorly with the unconstrained spectra, supporting the results of our cross validation test. Here, it should be noted that the BA spectra have been subject to broadening, and furthermore, the calculation of these spectra only take into consideration two-spinon and four-spinon contributions \cite{caux_05}. As such, it is known that a few percent of the total spectral weight is missing, which could at least partially explain the minor deviations between the two spectra in Fig.~\ref{hchain_spec}(c) and (f).

The fact that the edge outperforms the unconstrained and single-peak parameterization strongly supports the viability of cross validation as a tool for model selection. Considering the fact that for a chain of length $512$, the exact $T=0$ spectrum likely only contains $\sim$100 delta functions with significant weight, deduced by extrapolating from the results for smaller system sizes \cite{wang_19, xie_18}, it is remarkable that our method is able correctly identify the true form of the spectral function in the thermodynamic limit. It is quite possible that if smaller system size was considered instead, a spectrum composed of a dominate peak plus a smooth continuation would actually be the best representation and the single-peak parameterization would perform the best. But this is not the case here, implying that finite size effects are sufficiently mitigated.

To supplement these results, we have performed a second cross validation test in which the $G(\tau)$ bins are split into two mutually exclusive sets, instead of 21. Here, one set is used for sampling and the other is used for validation, but now we repeat the partitioning many times, preforming many runs of the cross validation procedure in a ``bootstrap-style.'' The validation $\chi^2$ value is then averaged over these repetitions and compared amongst the different SAC parameterizations. We performed this secondary test to ensure that the reduction of data quality, as a consequence of using only a fraction of the data for sampling, did not skew our results. The results agree with those shown in Fig.~\ref{hchain_anneal}, which we present and discuss in Appendix~\ref{half_sec}.


\subsection{Heisenberg Chain with Long-Range Interactions}\label{spec_param_2}
We now turn to a case where the exact features contained in the spectral function are unknown, the unfrustrated AFM Heisenberg chain with power-law decaying interactions \cite{yusuf_04, laflorencie_05}:
\begin{equation}
    H = \sum_{r=1}^{L/2} J_r \sum_{i=1}^L \boldsymbol{S}_i \cdot \boldsymbol{S_{i+r}},
\end{equation}
where
\begin{equation}
    J_r = G\frac{(-1)^{r-1}}{r^\alpha},\quad G = \left(1 + \sum_{r=2}^{L/2} \frac{1}{r^\alpha}\right)^{-1}.
\end{equation}
We note that the normalization of the coupling, $G$, chosen so that $\sum_r |J_r| = 1$, differs slightly from that used in previous related studies. Interest in this model is rooted in its possible connection to the 2D square lattice AFM Heisenberg model, as the long-range, staggered interactions allow for symmetry breaking and true long-range order to form in this 1D quantum magnet at $T = 0$, effectively increasing its dimensionality \cite{feiguin_21}. The excitation spectra of the long-range Heisenberg chain is a key ingredient in understanding the nature of its ground state, and the ability to resolve sharp features with the aforementioned constrained SAC sampling schemes may help clear up the long-standing uncertainties surrounding this model. An extensive study of this system will be published in a separate paper, where spectral functions are calculated in both the N\'{e}el ordered phase ($\alpha < \alpha_c = 2.22(1)$), as well as  in the quasi long-range ordered (QLRO) phase ($\alpha > \alpha_c $), on the other side of the quantum phase transition (QPT) \cite{yang_24}.

\begin{figure}[t]
    \centering
    \includegraphics[width = .9\columnwidth]{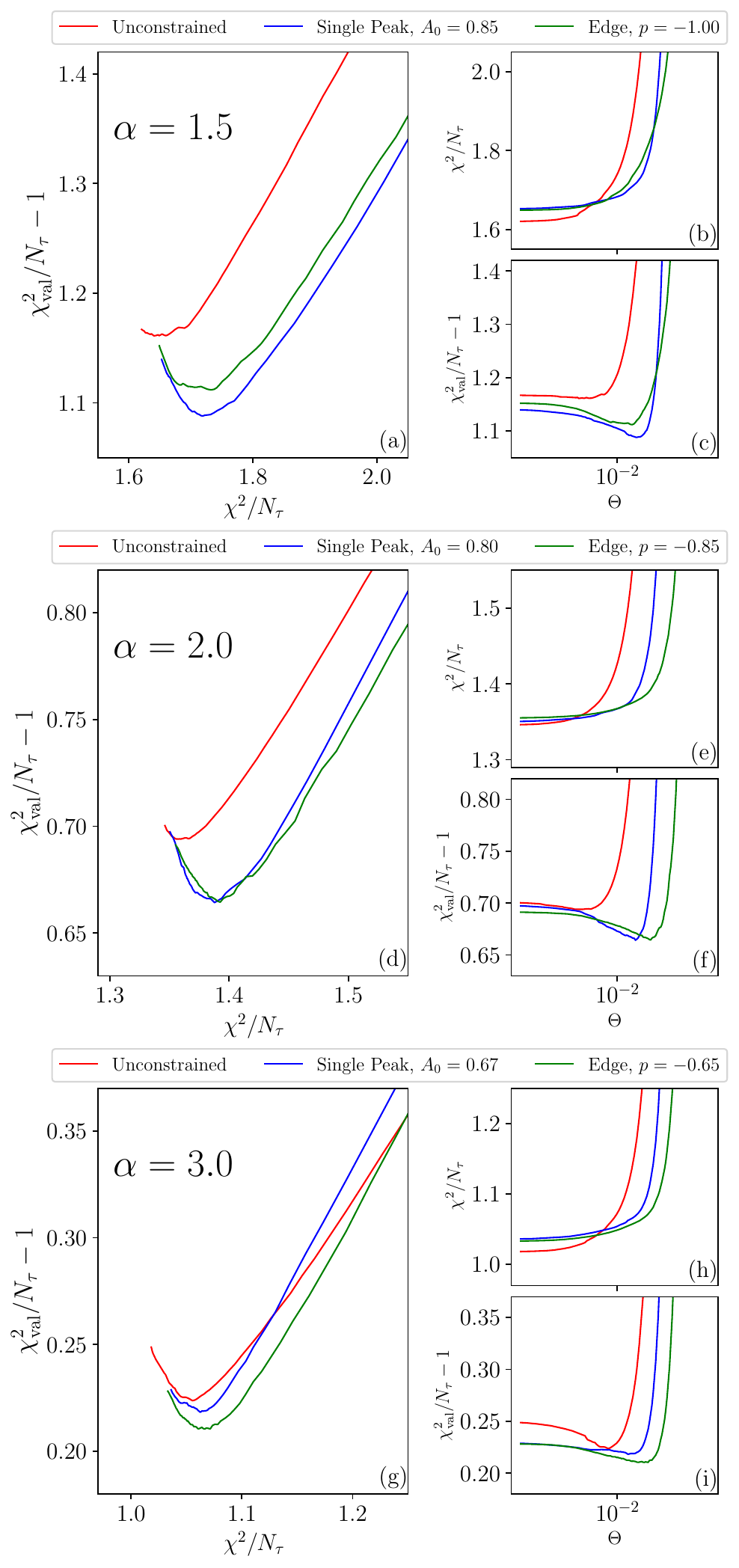}
    \caption{Cross validation results for the $S=1/2$ Heisenberg chain with long-range interactions ($L=256$, $\beta = 256$), for $\alpha  = 1.5$ (panels (a)-(c)), 2.0 (panels (d)-(f)), and 3.0 (panels (g)-(i)). The spectral function considered here is for the operator $O = S^z_{q=\pi/2}$. Panels and set up are the same as in Fig.~\ref{hchain_anneal}.}
    \label{lr_hchain_anneal}
\end{figure}

The transition between the N\'{e}el and QLRO phases provides an excellent application for cross validation, because of the uncertainty surrounding the nature of the ground state excitations. Deep in the QLRO regime, it is expected that the spectral functions contains a power-law edge, in light of the BA results discussed in Sec. \ref{spec_param_2}. While deep in the N\'{e}el phase, the spectral function should contain a dominant magnon-peak, in analogy to the 2D AFM Heisenberg model \cite{igarashi_92, wallin_93}. A previous study on a Heisenberg chain with similar long-range interactions only resolved spectral functions with sharp magnon peaks, followed by weak features at higher energy, when the system is in N\'{e}el phase \cite{feiguin_21}. This study used the time-dependent Density Matrix Renormalization Group (tDMRG) method, which is known to produce artificially broadened peaks due to limits on the width of the real-time window used in the calculations. Due to the uncertainties introduced by this method, the nature of the higher-energy continuum was not examined in detail. The $\delta$-function peak SAC parameterization could potentially provide a significant improvement to these results, and the application of the edge parameterization could provide completely new insights into this model.

\begin{figure}[t]
    \centering
    \includegraphics[width = .9\columnwidth]{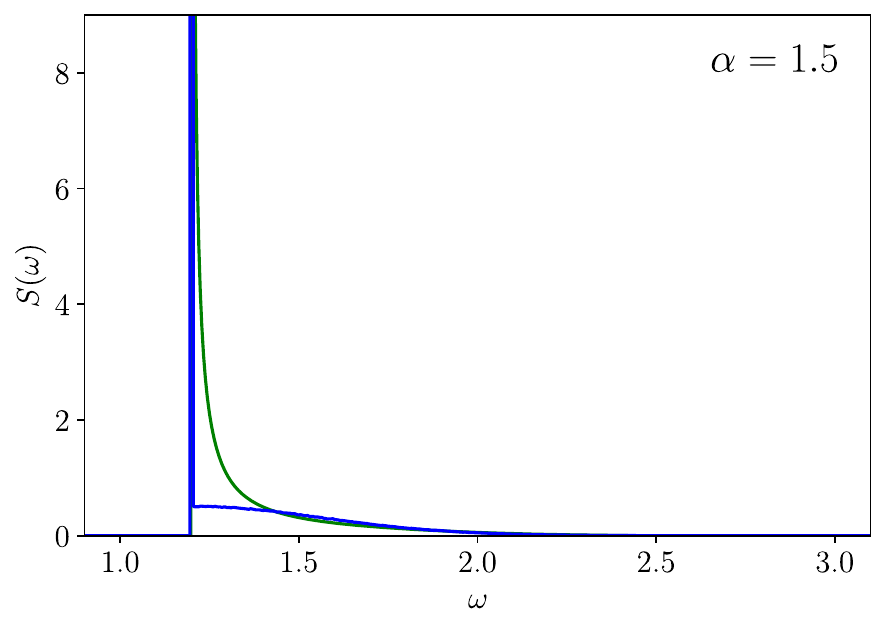}
    \caption{Comparison of the single-peak and edge parameterization spectral functions for the for the $S=1/2$ Heisenberg chain with long-range interactions, corresponding to the cross validation results shown in Fig.~\ref{hchain_anneal} panels (a)-(c). For the single-peak parameterization (blue) we use the leading $\delta$-function weight $A_0 = 0.85$ and $N_c = 2,000$ continuum $\delta$-functions. For the edge parameterization (green), we use $N_\omega = 640$ and the exponent $p = -1.00$.}
    \label{alpha15_spec}
\end{figure}

Here, we use cross validation to test three SAC parameterizations (unconstrained, single-peak, and edge) in the N\'{e}el ordered regime ($\alpha = 1.5$), in the N\'{e}el, but close to the QPT ($\alpha = 2.0$), and on the QLRO side of the QPT ($\alpha = 3.0$) for the spectral function of the operator $O = S^z_{q=\pi/2}$. In Ref.~\onlinecite{yang_24}, a broader set of SAC parameterizations are tested and all momenta are examined, but here, we will only focus on this single momentum and the three parameterizations discussed thus far.

Figure~\ref{lr_hchain_anneal} shows the results of the cross validation procedure for this model. Here, we again used QMC data generated using the SSE method, for a system of length $L=256$ at inverse temperature $\beta=256$. The number of validation data sets used in this run was $K=20$, which was chosen to limit the error level to $\sigma = 10^{-5}$. Just as before, the values of $A_0$ and $p$ used in these tests were identified using scans \cite{yang_24}.

For all values of $\alpha$ considered here, the unconstrained sampling expectedly performs the worst of the three parameterizations. It is known that in both the large and small $\alpha$ limits, the spectral function contains a sharp edge feature, so it is likely that for intermediate values of $\alpha$, the spectral function also contains a low energy edge of some form. It is most instructive, however, to look at the behavior of the single-peak and edge curves as $\alpha$ is increased. For $\alpha = 1.5$, the long-range Heisenberg chain exhibits N\'{e}el order and will thus host coherent, spin-wave excitations, albeit with anomalous dispersion \cite{yusuf_04, laflorencie_2005, yang_24}. We thus expect the spectral function to contain a dominant magnon peak, as was resolved using the single-peak parameterization in the 2D square lattice Heisenberg model \cite{sandvik_17}. Accordingly, the single-peak curve in Fig.~\ref{lr_hchain_anneal}(a) (blue) reaches the lowest validation $\chi^2$ minimum. The spectral functions produced using these two parameterizations for this value of $\alpha$ are shown in Fig.~\ref{alpha15_spec}. While the two spectral edges align nearly perfectly, the distributions of the spectral weight above this lower bound differ. We attribute the observed difference in the validation $\chi^2$ values to this difference and conclude that the spectral function for $\alpha = 1.5$ does not contain a monotonically decaying edge. 
We contrast that with the curves in Fig.~\ref{lr_hchain_anneal}(g). For $\alpha = 3.0$, the ground state of the long-range Heisenberg chain is much closer to that of the standard Heisenberg chain discussed in Sec.~\ref{spec_param}. This is reflected in the cross validation $\chi^2$, where the edge parameterization instead reaches the lowest minimum value.

Finally, for $\alpha = 2.0$ the validation $\chi^2$ does not appear to favor either of the single-peak or edge parameterizations due to the proximity to the QPT at $\alpha_c$. As $\alpha$ is increased from, the SAC parameterization that has the lowest validation $\chi^2$ switches over from the single-peak to the edge parameterization. This implies that there exists a crossover regime around the vicinity of $\alpha_c$ where cross validation is unable to identify a clear preference between the two spectral forms. The notion of a crossover between these two spectral forms is further supported by the smoothly increasing value of the optimal exponent $p$ as $\alpha$ is reduced. The edge parameterization spectra diverging more sharply  as the systems enters deeper into the N\'{e}el ordered phase can be considered a natural transition to the expected dominant $\delta$-peak form. However, in the intermediate regime near $\alpha_c$, it is unlikely that the spectral function can be described purely by one of these two sharp features. 

Different values of both $\alpha$ and momentum $q$ will be studied in Ref.~\onlinecite{yang_24}, but these preliminary tests provide further support for the viability of cross validation as a tool for model selection in the analytic continuation of QMC-generated data.

\section{Discussion}\label{conc}
We have explored the use of cross validation in the SAC method, as both a confirmation of the optimal $\Theta$ criterion, as well as a tool to select the most likely spectral parameterization for a given model. In our test cases using synthetic data, we found excellent agreement between the optimal $\Theta$ criterion Eq.~\eqref{criterion} and the location of the cross validation minimum. For the more complex of the two test cases, the double Gaussian peak spectrum, the broad validation $\chi^2$ minimum reflected the difficulties posed by the relatively sharp features contained in this spectrum. In principle this spectrum can be resolved using SAC, but the unconstrained sampling introduces entropic distortions that produce noticeable deviations from the exact result. As a result, all sampling temperatures within the broad minimum produce acceptable representations of the exact spectrum, which we demonstrated by using a $\Theta$ value one standard deviation below the average minimum value. Interestingly, at this lower sampling temperature some features of the spectrum were actually reproduced better. Since using cross validation to fix $\Theta$ would be a very poor use of computational resources, it is very encouraging that the optimal $\Theta$ criterion agrees so well with the the results of the validation procedure. 

A promising, and more decisive use of cross validation, is in model selection. The different entropy contents of the two unconstrained SAC parameterizations result in differing abilities to reproduce a given spectrum depending on its exact features. In some cases, it it advantageous to use the equal amplitudes parameterization, and in others the variable amplitudes parameterization performs substantially better. In previous studies, this was only able to be shown retrospectively, when comparing the SAC results to an artificial spectrum used to generate synthetic data. Here, we showed that cross validation can indeed act as a tool for model selection when comparing the two different unconstrained SAC parameterizations. This will be extremely valuable in practice, as often times these two parameterizations produce considerably different results, which can be even more dramatic when other parameterizations are considered, such as the fixed grid method \cite{shao_23} or the orthogonal polynomial representation \cite{wu_13}. When there is no artificial spectrum to compare to, it is impossible to determine which of these spectra is best.

Cross validation as a tool for model selection was also applied to the constrained SAC parameterizations. While the recent developments in the constrained sampling schemes have allowed for the resolution of spectra with sharp features, the ill-posed nature of the analytic continuation problem still presents obstacles for the implementation of these new methods. When many different spectra give acceptable $\chi^2$ values, cross validation is the least biased way to determine which parameterization is most likely, and our results for the $S=1/2$ AFM Heisenberg chain suggest that cross validation indeed can accomplish this task. Cross validation should be an extremely valuable tool in the study of highly correlated, many-body systems if it can be used to differentiate quantum phases of matter via their spectral features. Further tests on other systems with known spectral features will be very instructive.

The test on the Heisenberg chain with long-range interactions also produced promising results, correctly identify which side of the QPT the system was on based on the spectral features alone. While the exact features are only known with certainty in the limiting cases of $\alpha$, $\alpha\to 0$ and $\alpha\to \infty$, it is likely that these features persists throughout either side of the QPT. Further tests at more values of $\alpha$ and $q$ are still needed, and some will be presented in Ref.~\onlinecite{yang_24}, but as a general diagnostic tool, cross validation shows much promise in the practice of numerical analytic continuation of QMC-generated data.

Our cross validation scheme is not only applicable to SAC, but can also be applied other numerical analytic continuation methods, such as MEM. Based on our tests, a promising application of cross validation would be in the selection of the default model used in MEM. While not exactly analogous to the SAC parameterization, there are similarities between these two user-controlled inputs, such as their effects on the shape of the output spectra. Often times, perturbation theory is used to select the MEM default model, where some parameter in the perturbative solution is optimized to select the ``best'' spectra \cite{gubernatis_96}. Cross validation could be used to aide in this optimizations process or to compare entirely different default models, perhaps from different perturbative solutions of the model under consideration.

\begin{acknowledgements}
We would like to thank Markus Holzmann for useful discussions that inspired this study. This research was supported by the Simons Foundation under Grant No. 511064. The numerical calculations were carried out on the Shared Computing Cluster managed by Boston University’s Research Computing Services.
\end{acknowledgements}

\appendix

\section{Alternative Cross Validation Procedure}\label{half_sec}
As briefly introduced in Sec.~\ref{spec_param}, we have performed an alternative cross validation procedure that strays slightly from the scheme introduced in the main text. In this approach, we maintain as high a data quality as possible in both the sampling and validation data sets by splitting the $G(\tau)$ bins into two mutually exclusive sets, instead of 21. To reduce fluctuations in the $\chi^2_{\mathrm{val}}$ statistic, we repeat this process many times, with new random partitions for the sampling and validation sets, and average the results (500 repetitions for the examples we present here).

\begin{figure}[t]
\vspace{2mm}
\begin{center}
\includegraphics[width=0.95\columnwidth]{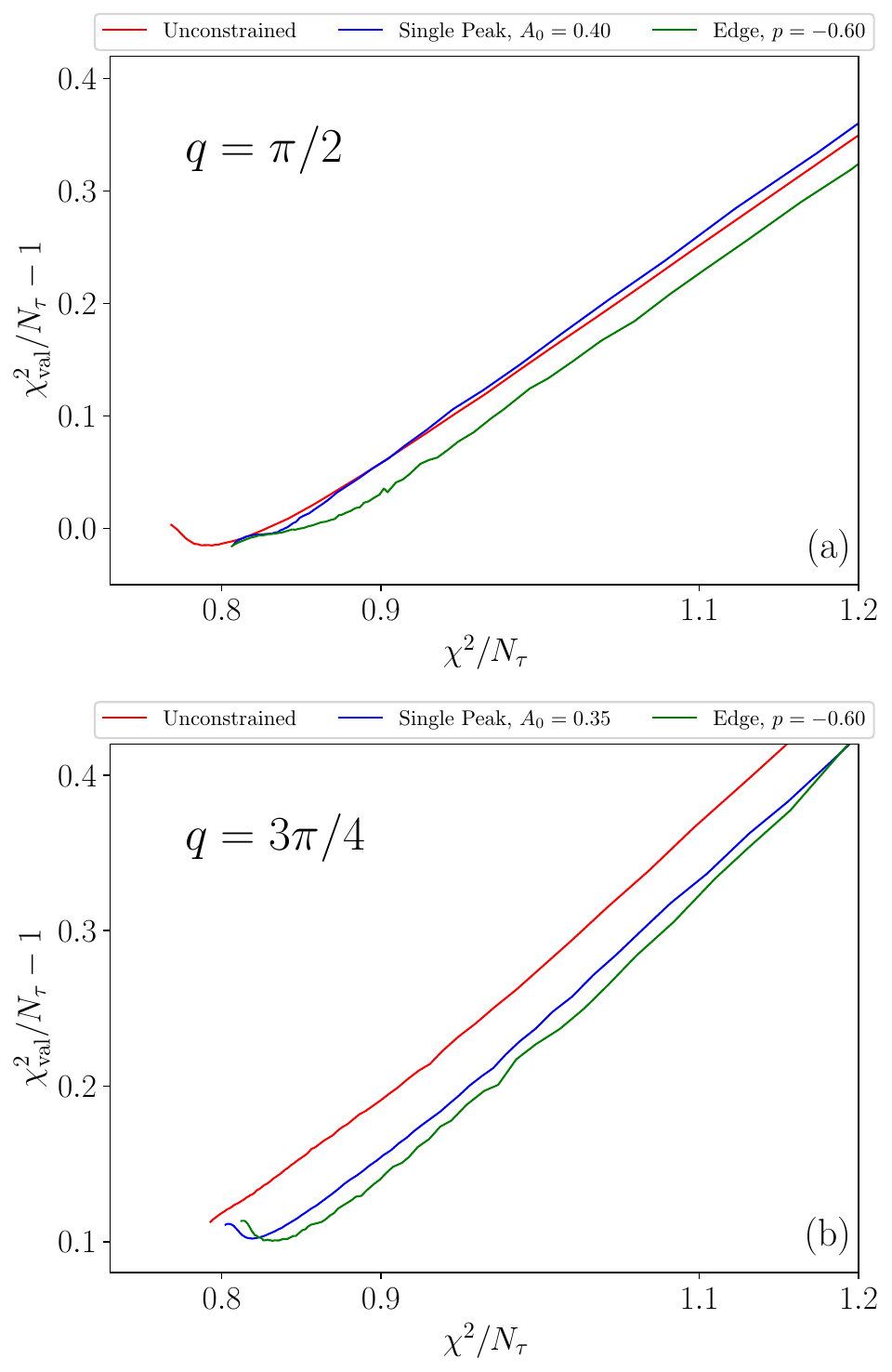}
\caption{Cross validation results for $S(q=\pi/2, \omega)$ and $S(q=3\pi/4, \omega)$ of the $S=1/2$ Heisenberg chain ($L=512$, $\beta = 1024$) using the alternative cross validation procedure. Panels (a) and (b) show validation $\chi^2$ as a function of the sampling $\chi^2$ for $q=\pi/2$ and $q=3\pi/4$, respectively.}
\label{alternative_cross_val}
\end{center}
\end{figure}

We tested this alternative procedure on the $S=1/2$ Heisenberg chain and present the results for for $S(q= \pi/2, \omega)$ and $S(q= 3\pi/4, \omega)$ in Fig.~\ref{alternative_cross_val}. The same conclusions are drawn from both cross validation methods (cf. Fig.~\ref{hchain_anneal}); the edge parameterization has the lowest validation $\chi^2$ when compared to the single-peak and unconstrained methods, at a fixed value of the $\chi^2$. The validation $\chi^2$ curves for the peak and edge parameterizations ($q=\pi/2$) and unconstrained sampling ($q=3\pi/4$) do not have well defined minima, which differs from the results in Fig.~\ref{hchain_anneal}. This, along with the the other intricacies involved in using this approach, will be studied more closely in a future paper. Nevertheless, the results of this test are clearly inline with those presented in Sec.~\ref{spec_param} and emphasize that the value of the validation $\chi^2$ throughout the entire anneal, as well as its minimum value, are important to consider when preforming cross validation.

This method offers several advantages, such as maintaining the highest possible data quality and the ability to preform many repetitions until the results have converged, but does not appear to produce results that conflict with our original approach. This is consistent with our experience using SAC. Increasing  the error level by a factor of $\sqrt{21} \approx 4.5$ (which is the effect of using only 1/21st of the total number of $G(\tau)$ bins) will rarely be detrimental to the accurately of the results, assuming the data quality is high to begin with and the spectral function does not contain many fine features. In a future study, we will investigate this balance of data quality and number of validation data sets in more detail, but these preliminary results suggest that lowered data quality has not impacted our cross validation procedure.

\section{Optimal SAC Parameters for the Heisenberg Chain}\label{appendix_scans}

\begin{figure}[t]
\vspace{2mm}
\begin{center}
\includegraphics[width=0.85\columnwidth]{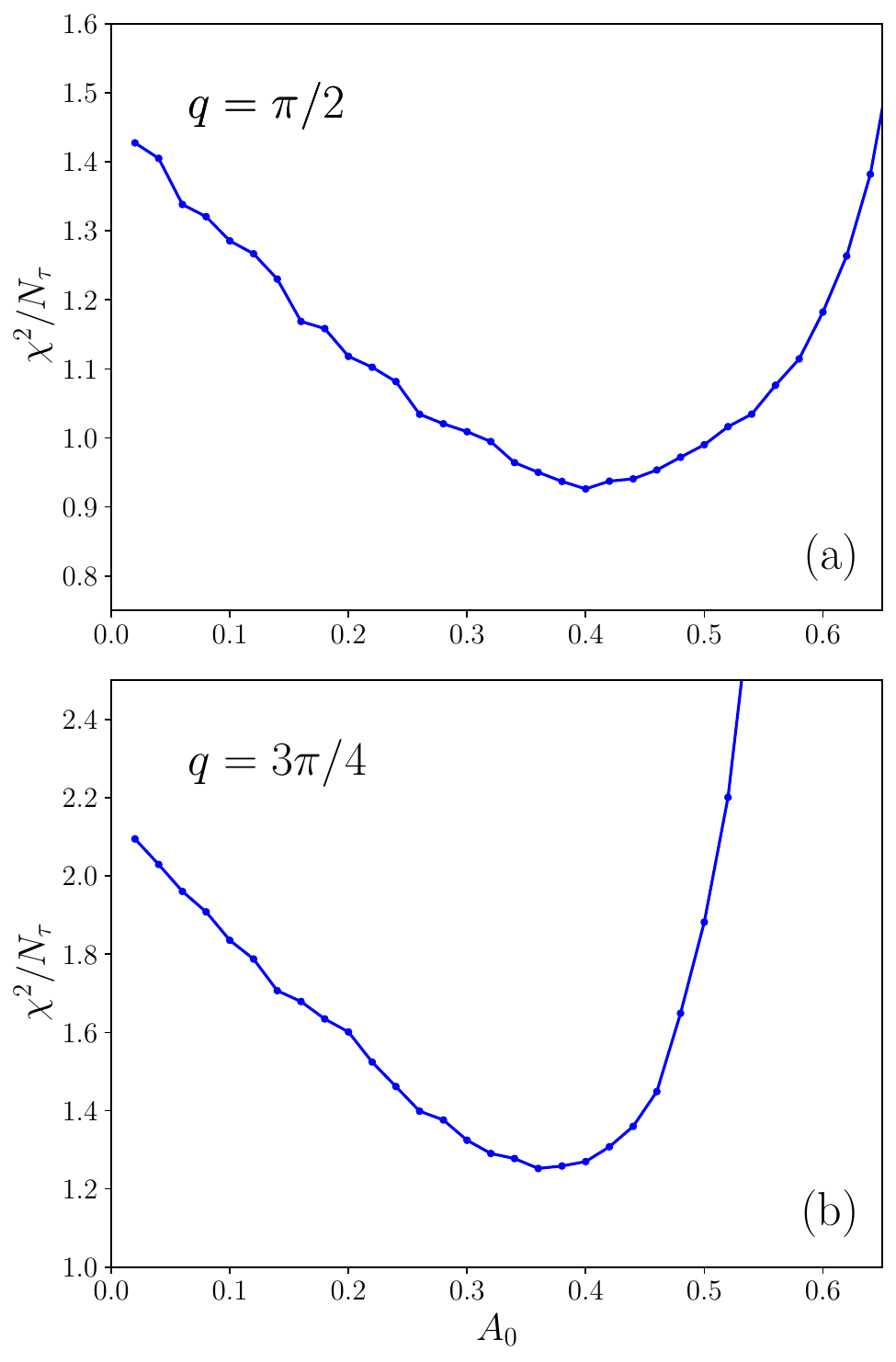}
\caption{Results of $A_0$ scan for the Heisenberg chain. Panels (a) and (b) show the $\chi^2$ value versus $A_0$ ($\Delta A_0 = 0.01$) for $q= \pi/2$ and $q= 3\pi/4$, respectively.}
\label{A0_scan_fig}
\end{center}
\end{figure}

When implementing the single-peak and edge SAC parameterizations, a fixed parameter must be chosen using a simple optimization procedure. In the case of the single-peak parameterization, this parameter is the weight of the macroscopic $\delta$-peak, $A_0$, and in the case of the edge parameterization, this parameter is the power $p$ with which the edge asymptotically diverges with. Scans over $A_0$ and $p$ are performed before cross validation to determine their optimal values.

In either case, the SAC annealing routine in run multiple times with different fixed values of $A_0$ or $p$, where typically the values are selected from a sensibly defined grid (not too dense or too sparse). After these scans are performed, the $\chi^2$ values as a function of either $A_0$ or $p$ are compared at a fixed value of the sampling temperature and a minimum is located. At very low values of $\Theta$, where the sampling is dominated by noise in the QMC-generated data, the $\chi^2$ minimum will be very shallow, if present at all, so instead an elevated temperature is used. This temperature can be chosen based on the minimum $\chi^2$ reach in an unrestricted sampling SAC run on the same set of data, as detailed in Ref.~\onlinecite{shao_23}, or simply set to a value high enough for a clear minimum to emerge (but low enough so that the $\chi^2$ value at the minimum is still acceptable). The exact location of the minimum depends on the sampling temperature to some extent, but in most cases this dependance is very minor. Furthermore, any value of $A_0$ or $p$ chosen around the $\chi^2$ minimum will produce nearly identical spectra.

It is crucial to have as highest possible data quality when determining the optimal values of $A_0$ and $p$. 
Therefore, we use the full set of QMC-generated data for the Heisenberg chain when performing the parameter scans. In this case, the complete $G(\tau)$ data set has an error level $\sim~3~\times~10^{-6}$, nearly ten times smaller than the error level of the validation data sets.

\subsection{$A_0$ Scan}\label{A0_scan}

Figure~\ref{A0_scan_fig} shows the results of the $A_0$ scans used to determine the optimal weight of the macroscopic $\delta$-peak for the two momenta considered, $q = \pi/2$, panel (a), and $q = 3\pi/4$, panel (b). In this case, unrestricted sampling runs (i.e. $A_0 = 0$) were first performed to determine the minimum $\chi^2$ values reached at the end of each anneal. This value is then used to determine the sampling temperature for the $A_0>0$ runs; we use a value of $\Theta$ that is slightly elevated from that corresponding to the criterion Eq.~\eqref{criterion} for this unrestricted sampling run \cite{shao_23}. At this temperature, a clear minimum is present, but the $\chi^2$ value is still at an acceptable value.
The optimal values of $A_0$, as determined by the results shown in Fig.~\ref{A0_scan_fig}, are $A_0=0.40$ and $0.35$ for $q=\pi/2$ and $q=3\pi/4$, respectively.

\begin{figure}[b]
\begin{center}
\vspace{2mm}
\includegraphics[width=0.9\columnwidth]{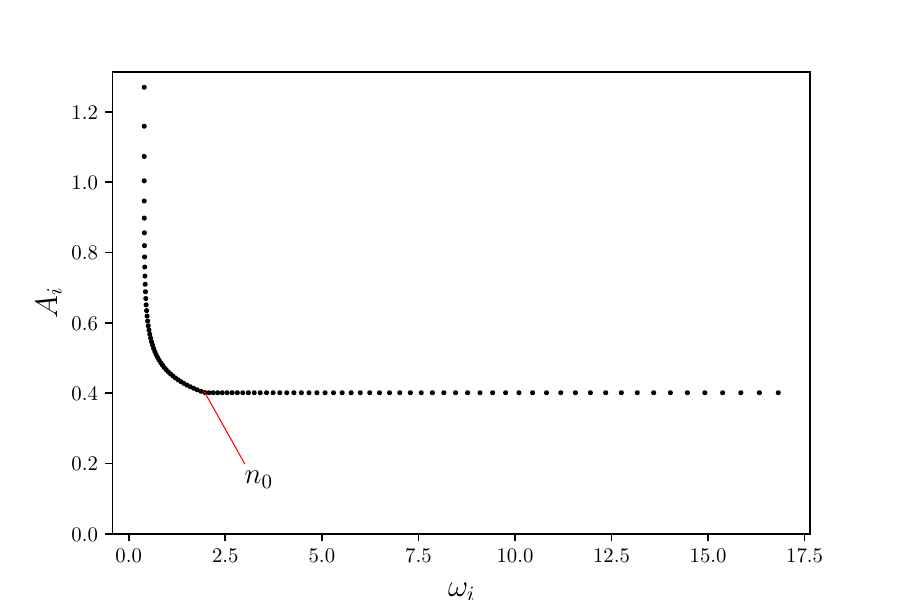}
\caption{Variable amplitudes used to produce the $\delta$-function diagram in Fig.~\ref{params}(d). For an exponent $p = -0.75$, the amplitudes increase as the edge is approached. The location of the equal amplitude cross-over point, $n_0$, is marked.}
\label{A_i_form_plot}
\end{center}
\end{figure}

\begin{figure}[t]
\begin{center}
\vspace{2mm}
\includegraphics[width=0.9\columnwidth]{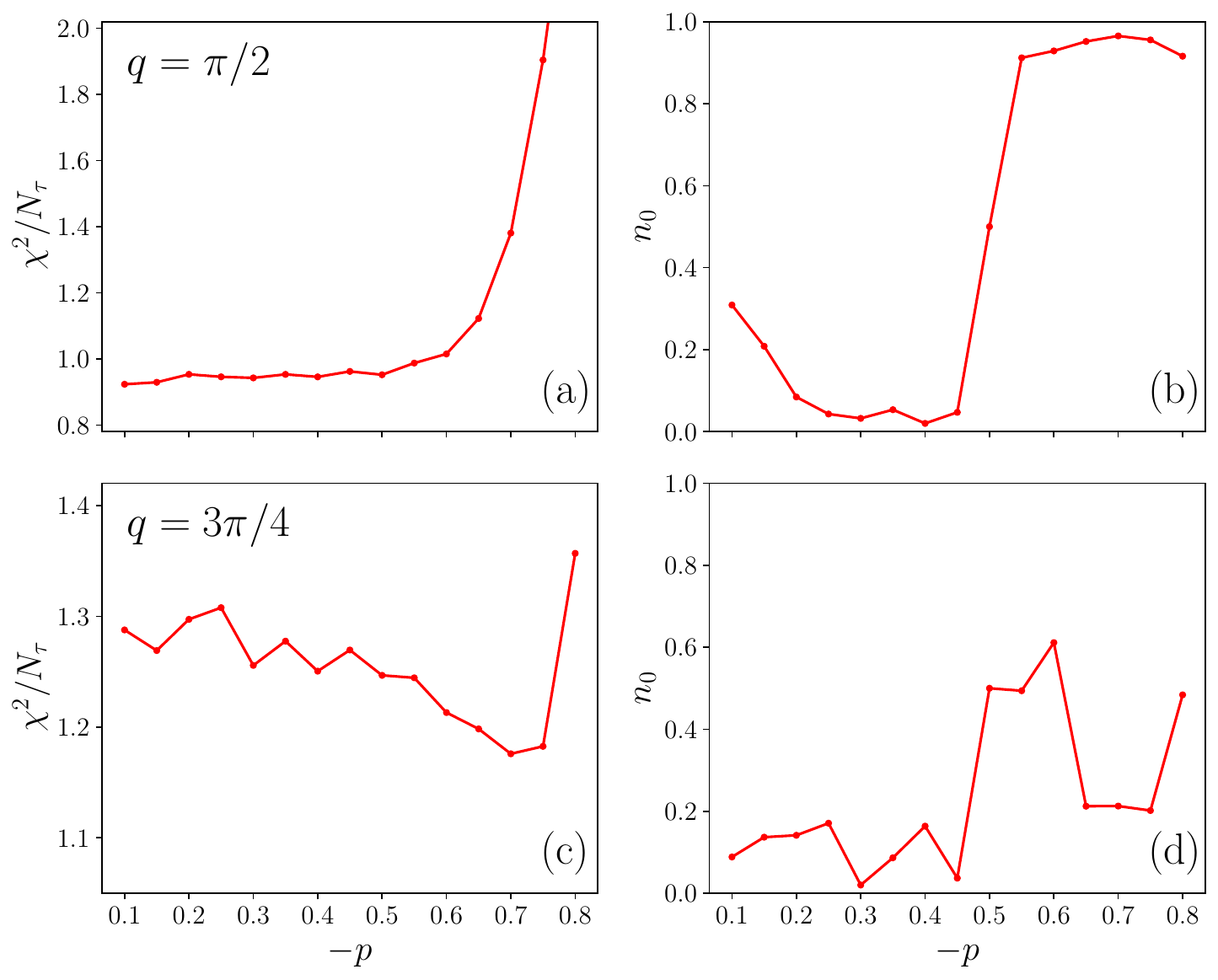}
\caption{Results of $p$ scan for the Heisenberg chain. Panels (a) and (c) show the $\chi^2$ value versus $p$ ($\Delta p = 0.05$) for $q= \pi/2$ and $q= 3\pi/4$, respectively and panels (b) and (d) show the corresponding values of $n_0$. }
\label{p_scan_fig}
\end{center}
\end{figure}

\subsection{$p$ Scan}\label{p_scan}

As discussed briefly in Sec.~\ref{spec_param}, and in detail in Ref.~\onlinecite{shao_23}, the edge parameterization is able to produce a spectrum with an edge that diverges asymptotically with an arbitrary exponent $p\neq -0.5$. This is achieved by varying the amplitudes of the sampled $\delta$-functions:
\begin{equation}\label{A_i_form}
	A_i \propto i^c, \;\;  c = 2p+1
\end{equation}
Here, the index $i \in \{1, N_\omega\}$ denotes the index of the sampled $\delta$-functions (note that $p = -0.5$ corresponds to $c = 0$, i.e. equal amplitudes). In both the $p=0.5$ and $p\neq -0.5$ cases, the spectrum will only exhibit the power-law divergent form very close to the edge; away from this edge, the QMC-generated data instead dictates the exact shape of the monotonically decaying tail. Therefore, it is not necessary to varying the amplitudes according to Eq.~\eqref{A_i_form} for all $N_\omega$ of the  $\delta$-functions, and the instead a cross-over to equal amplitudes is implemented at index $i = n_0$. The exact location of $n_0$ is determined by the QMC-generated data, and is thus sampled along with the other updates of the spectrum within the SAC program. Figure~\ref{A_i_form_plot} shows an example of how the amplitudes may vary with $\omega_i$ for an exponent $p=-3/4$, which is what was used for the for the diagram in Fig.~\ref{params}(d). The cross-over point for this example was $n_0/N_\omega = 0.50$, meaning that half of the amplitudes vary according to Eq.~\eqref{A_i_form}.

Because of this, one has to be slightly more careful when analyzing the results of a scan over $p$. If the $p$ value that minimizes the $\chi^2$ value is accompanied by a very small value of $n_0$, this implies that this optimal exponent is ``fictitious.'' In this case, the $\delta$-functions transition very quickly to the equal amplitude form and the asymptotic divergence $(\omega-\omega_0)^{p}$ is only realized is a very small region of the spectrum. In contrast, if the optimal exponent has a $n_0$ value that is close to one, the amplitude form is applied to the majority of the $\delta$-functions and is thus realized for a large portion of the spectrum. This suggests that the accompanying low $\chi^2$ value indeed indicates a good statistical fit for a spectrum with this specific divergent edge form.

There are no definite rules for determining the optimal exponent under these considerations, but Fig.~\ref{p_scan_fig} provides two characteristic examples of how one may use the $\chi^2$ and $n_0$ values from a scan over $p$ to determine the optimal exponent. Panels (a) and (b) show the $\chi^2$ and $n_0$ values versus $-p$ for $S(q=\pi/2, \omega)$. In this case, the $\chi^2$ value is nearly constant for $-p< 0.50$ and rapidly increases  when $-p$ exceeds 0.50. The value of $n_0$ is small (less than $\sim 0.30$ and as small as 0.02 for $-p=0.40$) during this constant $\chi^2$ platform, which explains this behavior. The low values of  $n_0$ imply that the edge is, for the most part, of the form $(\omega-\omega_0)^{-1/2}$, which is corroborated by the fact that the $\chi^2$ value when $p$ is exactly $-0.50$ is roughly equal to the $\chi^2$ values for $-p< 0.50$.

When $-p$ exceeds 0.50, the value of $n_0$ jumps close to one. Since the $\chi^2$ value also begins to increase in this region, $-p>0.50$ can be ruled out as optimal exponent values. The combination of these two trends strongly imply that $p=-0.50$ is the true exponent, which is to be fully expected for dynamic spin structure factor of the Heisenberg chain, calculated using BA solution \cite{maillet_05, caux_2005, affleck_06}. 

Panels (c) and (d) show the results of the $p$ scan for $S(q=3\pi/4, \omega)$. Just as for $q=\pi/2$, $n_0$ is small for $-p<0.50$ (less than $\sim 0.20$ and as small as 0.02 for $-p=0.30$). However, in contrast to $q=\pi/2$, this range of $p$ values does not coincide with the $\chi^2$ minimum. Therefore, we can say with certainty that the optimal exponent does not lie within this range. In this case, there is a clear $\chi^2$ minimum at $-p=0.70$, but the accompanying value of $n_0$ is small, $\sim 0.20$. The $n_0$ value for $-p=0.60$ is much larger, around $0.60$, and the $\chi^2$ value is within 4 \% of the overall minimum value. This difference in the $\chi^2$ is likely too small to say with certainty that $p=-0.70$ is really preferable over $p=-0.60$, so considering the much larger $n_0$ value, we conclude that the true optimal exponent for $q=3\pi/4$ is $p = -0.60$. This is further supported by the very good agreement between the SAC spectra with $p = -0.60$ and the BA solution, shown in Fig.~\ref{hchain_spec}(f).


\vfill

\bibliography{bib.bib}

\begin{thebibliography}{41}%
\makeatletter
\providecommand \@ifxundefined [1]{%
 \@ifx{#1\undefined}
}%
\providecommand \@ifnum [1]{%
 \ifnum #1\expandafter \@firstoftwo
 \else \expandafter \@secondoftwo
 \fi
}%
\providecommand \@ifx [1]{%
 \ifx #1\expandafter \@firstoftwo
 \else \expandafter \@secondoftwo
 \fi
}%
\providecommand \natexlab [1]{#1}%
\providecommand \enquote  [1]{``#1''}%
\providecommand \bibnamefont  [1]{#1}%
\providecommand \bibfnamefont [1]{#1}%
\providecommand \citenamefont [1]{#1}%
\providecommand \href@noop [0]{\@secondoftwo}%
\providecommand \href [0]{\begingroup \@sanitize@url \@href}%
\providecommand \@href[1]{\@@startlink{#1}\@@href}%
\providecommand \@@href[1]{\endgroup#1\@@endlink}%
\providecommand \@sanitize@url [0]{\catcode `\\12\catcode `\$12\catcode
  `\&12\catcode `\#12\catcode `\^12\catcode `\_12\catcode `\%12\relax}%
\providecommand \@@startlink[1]{}%
\providecommand \@@endlink[0]{}%
\providecommand \url  [0]{\begingroup\@sanitize@url \@url }%
\providecommand \@url [1]{\endgroup\@href {#1}{\urlprefix }}%
\providecommand \urlprefix  [0]{URL }%
\providecommand \Eprint [0]{\href }%
\providecommand \doibase [0]{https://doi.org/}%
\providecommand \selectlanguage [0]{\@gobble}%
\providecommand \bibinfo  [0]{\@secondoftwo}%
\providecommand \bibfield  [0]{\@secondoftwo}%
\providecommand \translation [1]{[#1]}%
\providecommand \BibitemOpen [0]{}%
\providecommand \bibitemStop [0]{}%
\providecommand \bibitemNoStop [0]{.\EOS\space}%
\providecommand \EOS [0]{\spacefactor3000\relax}%
\providecommand \BibitemShut  [1]{\csname bibitem#1\endcsname}%
\let\auto@bib@innerbib\@empty
\bibitem [{\citenamefont {Sch\"uttler}\ and\ \citenamefont
  {Scalapino}(1985)}]{schuttler_85}%
  \BibitemOpen
  \bibfield  {author} {\bibinfo {author} {\bibfnamefont {H.~B.}\ \bibnamefont
  {Sch\"uttler}}\ and\ \bibinfo {author} {\bibfnamefont {D.~J.}\ \bibnamefont
  {Scalapino}},\ }\bibfield  {title} {\bibinfo {title} {Monte carlo studies of
  the dynamics of quantum many-body systems},\ }\href
  {https://doi.org/10.1103/PhysRevLett.55.1204} {\bibfield  {journal} {\bibinfo
   {journal} {Phys. Rev. Lett.}\ }\textbf {\bibinfo {volume} {55}},\ \bibinfo
  {pages} {1204} (\bibinfo {year} {1985})}\BibitemShut {NoStop}%
\bibitem [{\citenamefont {White}\ \emph {et~al.}(1989)\citenamefont {White},
  \citenamefont {Scalapino}, \citenamefont {Sugar},\ and\ \citenamefont
  {Bickers}}]{white_89}%
  \BibitemOpen
  \bibfield  {author} {\bibinfo {author} {\bibfnamefont {S.~R.}\ \bibnamefont
  {White}}, \bibinfo {author} {\bibfnamefont {D.~J.}\ \bibnamefont
  {Scalapino}}, \bibinfo {author} {\bibfnamefont {R.~L.}\ \bibnamefont
  {Sugar}},\ and\ \bibinfo {author} {\bibfnamefont {N.~E.}\ \bibnamefont
  {Bickers}},\ }\bibfield  {title} {\bibinfo {title} {Monte carlo calculation
  of dynamical properties of the two-dimensional hubbard model},\ }\href
  {https://doi.org/10.1103/PhysRevLett.63.1523} {\bibfield  {journal} {\bibinfo
   {journal} {Phys. Rev. Lett.}\ }\textbf {\bibinfo {volume} {63}},\ \bibinfo
  {pages} {1523} (\bibinfo {year} {1989})}\BibitemShut {NoStop}%
\bibitem [{\citenamefont {Jarrell}\ and\ \citenamefont
  {Biham}(1989)}]{jarrell_89}%
  \BibitemOpen
  \bibfield  {author} {\bibinfo {author} {\bibfnamefont {M.}~\bibnamefont
  {Jarrell}}\ and\ \bibinfo {author} {\bibfnamefont {O.}~\bibnamefont
  {Biham}},\ }\bibfield  {title} {\bibinfo {title} {Dynamical approach to
  analytic continuation of quantum monte carlo data},\ }\href
  {https://doi.org/10.1103/PhysRevLett.63.2504} {\bibfield  {journal} {\bibinfo
   {journal} {Phys. Rev. Lett.}\ }\textbf {\bibinfo {volume} {63}},\ \bibinfo
  {pages} {2504} (\bibinfo {year} {1989})}\BibitemShut {NoStop}%
\bibitem [{\citenamefont {Silver}\ \emph
  {et~al.}(1990{\natexlab{a}})\citenamefont {Silver}, \citenamefont {Sivia},\
  and\ \citenamefont {Gubernatis}}]{silver_90_1}%
  \BibitemOpen
  \bibfield  {author} {\bibinfo {author} {\bibfnamefont {R.~N.}\ \bibnamefont
  {Silver}}, \bibinfo {author} {\bibfnamefont {D.~S.}\ \bibnamefont {Sivia}},\
  and\ \bibinfo {author} {\bibfnamefont {J.~E.}\ \bibnamefont {Gubernatis}},\
  }\bibfield  {title} {\bibinfo {title} {Maximum-entropy method for analytic
  continuation of quantum monte carlo data},\ }\href
  {https://doi.org/10.1103/PhysRevB.41.2380} {\bibfield  {journal} {\bibinfo
  {journal} {Phys. Rev. B}\ }\textbf {\bibinfo {volume} {41}},\ \bibinfo
  {pages} {2380} (\bibinfo {year} {1990}{\natexlab{a}})}\BibitemShut {NoStop}%
\bibitem [{\citenamefont {Silver}\ \emph
  {et~al.}(1990{\natexlab{b}})\citenamefont {Silver}, \citenamefont
  {Gubernatis}, \citenamefont {Sivia},\ and\ \citenamefont
  {Jarrell}}]{silver_90_2}%
  \BibitemOpen
  \bibfield  {author} {\bibinfo {author} {\bibfnamefont {R.~N.}\ \bibnamefont
  {Silver}}, \bibinfo {author} {\bibfnamefont {J.~E.}\ \bibnamefont
  {Gubernatis}}, \bibinfo {author} {\bibfnamefont {D.~S.}\ \bibnamefont
  {Sivia}},\ and\ \bibinfo {author} {\bibfnamefont {M.}~\bibnamefont
  {Jarrell}},\ }\bibfield  {title} {\bibinfo {title} {Spectral densities of the
  symmetric anderson model},\ }\href
  {https://doi.org/10.1103/PhysRevLett.65.496} {\bibfield  {journal} {\bibinfo
  {journal} {Phys. Rev. Lett.}\ }\textbf {\bibinfo {volume} {65}},\ \bibinfo
  {pages} {496} (\bibinfo {year} {1990}{\natexlab{b}})}\BibitemShut {NoStop}%
\bibitem [{\citenamefont {Gubernatis}\ \emph {et~al.}(1991)\citenamefont
  {Gubernatis}, \citenamefont {Jarrell}, \citenamefont {Silver},\ and\
  \citenamefont {Sivia}}]{gubernatis_91}%
  \BibitemOpen
  \bibfield  {author} {\bibinfo {author} {\bibfnamefont {J.~E.}\ \bibnamefont
  {Gubernatis}}, \bibinfo {author} {\bibfnamefont {M.}~\bibnamefont {Jarrell}},
  \bibinfo {author} {\bibfnamefont {R.~N.}\ \bibnamefont {Silver}},\ and\
  \bibinfo {author} {\bibfnamefont {D.~S.}\ \bibnamefont {Sivia}},\ }\bibfield
  {title} {\bibinfo {title} {Quantum monte carlo simulations and maximum
  entropy: Dynamics from imaginary-time data},\ }\href
  {https://doi.org/10.1103/PhysRevB.44.6011} {\bibfield  {journal} {\bibinfo
  {journal} {Phys. Rev. B}\ }\textbf {\bibinfo {volume} {44}},\ \bibinfo
  {pages} {6011} (\bibinfo {year} {1991})}\BibitemShut {NoStop}%
\bibitem [{\citenamefont {Jarrell}\ and\ \citenamefont
  {Gubernatis}(1996)}]{gubernatis_96}%
  \BibitemOpen
  \bibfield  {author} {\bibinfo {author} {\bibfnamefont {M.}~\bibnamefont
  {Jarrell}}\ and\ \bibinfo {author} {\bibfnamefont {J.}~\bibnamefont
  {Gubernatis}},\ }\bibfield  {title} {\bibinfo {title} {Bayesian inference and
  the analytic continuation of imaginary-time quantum monte carlo data},\
  }\href {https://doi.org/https://doi.org/10.1016/0370-1573(95)00074-7}
  {\bibfield  {journal} {\bibinfo  {journal} {Physics Reports}\ }\textbf
  {\bibinfo {volume} {269}},\ \bibinfo {pages} {133} (\bibinfo {year}
  {1996})}\BibitemShut {NoStop}%
\bibitem [{\citenamefont {Bergeron}\ and\ \citenamefont
  {Tremblay}(2016)}]{bergeron_16}%
  \BibitemOpen
  \bibfield  {author} {\bibinfo {author} {\bibfnamefont {D.}~\bibnamefont
  {Bergeron}}\ and\ \bibinfo {author} {\bibfnamefont {A.-M.~S.}\ \bibnamefont
  {Tremblay}},\ }\bibfield  {title} {\bibinfo {title} {Algorithms for optimized
  maximum entropy and diagnostic tools for analytic continuation},\ }\href
  {https://doi.org/10.1103/PhysRevE.94.023303} {\bibfield  {journal} {\bibinfo
  {journal} {Phys. Rev. E}\ }\textbf {\bibinfo {volume} {94}},\ \bibinfo
  {pages} {023303} (\bibinfo {year} {2016})}\BibitemShut {NoStop}%
\bibitem [{\citenamefont {White}(1991)}]{white_91}%
  \BibitemOpen
  \bibfield  {author} {\bibinfo {author} {\bibfnamefont {S.~R.}\ \bibnamefont
  {White}},\ }\bibfield  {title} {\bibinfo {title} {The average spectrum method
  for the analytic continuation of imaginary-time data},\ }in\ \href@noop {}
  {\emph {\bibinfo {booktitle} {Computer Simulation Studies in Condensed Matter
  Physics III}}},\ \bibinfo {editor} {edited by\ \bibinfo {editor}
  {\bibfnamefont {D.~P.}\ \bibnamefont {Landau}}, \bibinfo {editor}
  {\bibfnamefont {K.~K.}\ \bibnamefont {Mon}},\ and\ \bibinfo {editor}
  {\bibfnamefont {H.-B.}\ \bibnamefont {Sch{\"u}ttler}}}\ (\bibinfo
  {publisher} {Springer Berlin Heidelberg},\ \bibinfo {address} {Berlin,
  Heidelberg},\ \bibinfo {year} {1991})\ pp.\ \bibinfo {pages}
  {145--153}\BibitemShut {NoStop}%
\bibitem [{\citenamefont {Sandvik}(1998)}]{sandvik_98}%
  \BibitemOpen
  \bibfield  {author} {\bibinfo {author} {\bibfnamefont {A.~W.}\ \bibnamefont
  {Sandvik}},\ }\bibfield  {title} {\bibinfo {title} {Stochastic method for
  analytic continuation of quantum monte carlo data},\ }\href
  {https://doi.org/10.1103/PhysRevB.57.10287} {\bibfield  {journal} {\bibinfo
  {journal} {Phys. Rev. B}\ }\textbf {\bibinfo {volume} {57}},\ \bibinfo
  {pages} {10287} (\bibinfo {year} {1998})}\BibitemShut {NoStop}%
\bibitem [{\citenamefont {Beach}(2004)}]{beach_04}%
  \BibitemOpen
  \bibfield  {author} {\bibinfo {author} {\bibfnamefont {K.~S.~D.}\
  \bibnamefont {Beach}},\ }\href@noop {} {\bibinfo {title} {Identifying the
  maximum entropy method as a special limit of stochastic analytic
  continuation}} (\bibinfo {year} {2004}),\ \Eprint
  {https://arxiv.org/abs/cond-mat/0403055} {arXiv:cond-mat/0403055
  [cond-mat.str-el]} \BibitemShut {NoStop}%
\bibitem [{\citenamefont {Vafayi}\ and\ \citenamefont
  {Gunnarsson}(2007)}]{vafayi_07}%
  \BibitemOpen
  \bibfield  {author} {\bibinfo {author} {\bibfnamefont {K.}~\bibnamefont
  {Vafayi}}\ and\ \bibinfo {author} {\bibfnamefont {O.}~\bibnamefont
  {Gunnarsson}},\ }\bibfield  {title} {\bibinfo {title} {Analytical
  continuation of spectral data from imaginary time axis to real frequency axis
  using statistical sampling},\ }\href
  {https://doi.org/10.1103/PhysRevB.76.035115} {\bibfield  {journal} {\bibinfo
  {journal} {Phys. Rev. B}\ }\textbf {\bibinfo {volume} {76}},\ \bibinfo
  {pages} {035115} (\bibinfo {year} {2007})}\BibitemShut {NoStop}%
\bibitem [{\citenamefont {Reichman}\ and\ \citenamefont
  {Rabani}(2009)}]{reichman_09}%
  \BibitemOpen
  \bibfield  {author} {\bibinfo {author} {\bibfnamefont {D.~R.}\ \bibnamefont
  {Reichman}}\ and\ \bibinfo {author} {\bibfnamefont {E.}~\bibnamefont
  {Rabani}},\ }\bibfield  {title} {\bibinfo {title} {{Analytic continuation
  average spectrum method for quantum liquids}},\ }\href
  {https://doi.org/10.1063/1.3185728} {\bibfield  {journal} {\bibinfo
  {journal} {The Journal of Chemical Physics}\ }\textbf {\bibinfo {volume}
  {131}},\ \bibinfo {pages} {054502} (\bibinfo {year} {2009})}\BibitemShut
  {NoStop}%
\bibitem [{\citenamefont {Sylju\aa{}sen}(2008)}]{olav_08}%
  \BibitemOpen
  \bibfield  {author} {\bibinfo {author} {\bibfnamefont {O.~F.}\ \bibnamefont
  {Sylju\aa{}sen}},\ }\bibfield  {title} {\bibinfo {title} {Using the average
  spectrum method to extract dynamics from quantum monte carlo simulations},\
  }\href {https://doi.org/10.1103/PhysRevB.78.174429} {\bibfield  {journal}
  {\bibinfo  {journal} {Phys. Rev. B}\ }\textbf {\bibinfo {volume} {78}},\
  \bibinfo {pages} {174429} (\bibinfo {year} {2008})}\BibitemShut {NoStop}%
\bibitem [{\citenamefont {Fuchs}\ \emph {et~al.}(2010)\citenamefont {Fuchs},
  \citenamefont {Pruschke},\ and\ \citenamefont {Jarrell}}]{fuchs_10}%
  \BibitemOpen
  \bibfield  {author} {\bibinfo {author} {\bibfnamefont {S.}~\bibnamefont
  {Fuchs}}, \bibinfo {author} {\bibfnamefont {T.}~\bibnamefont {Pruschke}},\
  and\ \bibinfo {author} {\bibfnamefont {M.}~\bibnamefont {Jarrell}},\
  }\bibfield  {title} {\bibinfo {title} {Analytic continuation of quantum monte
  carlo data by stochastic analytical inference},\ }\href
  {https://doi.org/10.1103/PhysRevE.81.056701} {\bibfield  {journal} {\bibinfo
  {journal} {Phys. Rev. E}\ }\textbf {\bibinfo {volume} {81}},\ \bibinfo
  {pages} {056701} (\bibinfo {year} {2010})}\BibitemShut {NoStop}%
\bibitem [{\citenamefont {Sandvik}(2016)}]{sandvik_16}%
  \BibitemOpen
  \bibfield  {author} {\bibinfo {author} {\bibfnamefont {A.~W.}\ \bibnamefont
  {Sandvik}},\ }\bibfield  {title} {\bibinfo {title} {Constrained sampling
  method for analytic continuation},\ }\href
  {https://doi.org/10.1103/PhysRevE.94.063308} {\bibfield  {journal} {\bibinfo
  {journal} {Phys. Rev. E}\ }\textbf {\bibinfo {volume} {94}},\ \bibinfo
  {pages} {063308} (\bibinfo {year} {2016})}\BibitemShut {NoStop}%
\bibitem [{\citenamefont {Qin}\ \emph {et~al.}(2017)\citenamefont {Qin},
  \citenamefont {Normand}, \citenamefont {Sandvik},\ and\ \citenamefont
  {Meng}}]{qin_17}%
  \BibitemOpen
  \bibfield  {author} {\bibinfo {author} {\bibfnamefont {Y.~Q.}\ \bibnamefont
  {Qin}}, \bibinfo {author} {\bibfnamefont {B.}~\bibnamefont {Normand}},
  \bibinfo {author} {\bibfnamefont {A.~W.}\ \bibnamefont {Sandvik}},\ and\
  \bibinfo {author} {\bibfnamefont {Z.~Y.}\ \bibnamefont {Meng}},\ }\bibfield
  {title} {\bibinfo {title} {Amplitude mode in three-dimensional dimerized
  antiferromagnets},\ }\href {https://doi.org/10.1103/PhysRevLett.118.147207}
  {\bibfield  {journal} {\bibinfo  {journal} {Phys. Rev. Lett.}\ }\textbf
  {\bibinfo {volume} {118}},\ \bibinfo {pages} {147207} (\bibinfo {year}
  {2017})}\BibitemShut {NoStop}%
\bibitem [{\citenamefont {Ghanem}\ and\ \citenamefont
  {Koch}(2020{\natexlab{a}})}]{ghanem_20}%
  \BibitemOpen
  \bibfield  {author} {\bibinfo {author} {\bibfnamefont {K.}~\bibnamefont
  {Ghanem}}\ and\ \bibinfo {author} {\bibfnamefont {E.}~\bibnamefont {Koch}},\
  }\bibfield  {title} {\bibinfo {title} {Average spectrum method for analytic
  continuation: Efficient blocked-mode sampling and dependence on the
  discretization grid},\ }\href {https://doi.org/10.1103/PhysRevB.101.085111}
  {\bibfield  {journal} {\bibinfo  {journal} {Phys. Rev. B}\ }\textbf {\bibinfo
  {volume} {101}},\ \bibinfo {pages} {085111} (\bibinfo {year}
  {2020}{\natexlab{a}})}\BibitemShut {NoStop}%
\bibitem [{\citenamefont {Ghanem}\ and\ \citenamefont
  {Koch}(2020{\natexlab{b}})}]{koch_20}%
  \BibitemOpen
  \bibfield  {author} {\bibinfo {author} {\bibfnamefont {K.}~\bibnamefont
  {Ghanem}}\ and\ \bibinfo {author} {\bibfnamefont {E.}~\bibnamefont {Koch}},\
  }\bibfield  {title} {\bibinfo {title} {Extending the average spectrum method:
  Grid point sampling and density averaging},\ }\href
  {https://doi.org/10.1103/PhysRevB.102.035114} {\bibfield  {journal} {\bibinfo
   {journal} {Phys. Rev. B}\ }\textbf {\bibinfo {volume} {102}},\ \bibinfo
  {pages} {035114} (\bibinfo {year} {2020}{\natexlab{b}})}\BibitemShut
  {NoStop}%
\bibitem [{\citenamefont {Shao}\ and\ \citenamefont {Sandvik}(2023)}]{shao_23}%
  \BibitemOpen
  \bibfield  {author} {\bibinfo {author} {\bibfnamefont {H.}~\bibnamefont
  {Shao}}\ and\ \bibinfo {author} {\bibfnamefont {A.~W.}\ \bibnamefont
  {Sandvik}},\ }\bibfield  {title} {\bibinfo {title} {Progress on stochastic
  analytic continuation of quantum monte carlo data},\ }\href
  {https://doi.org/https://doi.org/10.1016/j.physrep.2022.11.002} {\bibfield
  {journal} {\bibinfo  {journal} {Phys. Rep.}\ }\textbf {\bibinfo {volume}
  {1003}},\ \bibinfo {pages} {1} (\bibinfo {year} {2023})},\ \bibinfo {note}
  {progress on stochastic analytic continuation of quantum Monte Carlo
  data}\BibitemShut {NoStop}%
\bibitem [{\citenamefont {Ghanem}\ and\ \citenamefont
  {Koch}(2023)}]{ghanem_23}%
  \BibitemOpen
  \bibfield  {author} {\bibinfo {author} {\bibfnamefont {K.}~\bibnamefont
  {Ghanem}}\ and\ \bibinfo {author} {\bibfnamefont {E.}~\bibnamefont {Koch}},\
  }\bibfield  {title} {\bibinfo {title} {Generalized maximum entropy methods as
  limits of the average spectrum method},\ }\href
  {https://doi.org/10.1103/PhysRevB.108.L201107} {\bibfield  {journal}
  {\bibinfo  {journal} {Phys. Rev. B}\ }\textbf {\bibinfo {volume} {108}},\
  \bibinfo {pages} {L201107} (\bibinfo {year} {2023})}\BibitemShut {NoStop}%
\bibitem [{\citenamefont {Bishop}(2006)}]{bishop_06}%
  \BibitemOpen
  \bibfield  {author} {\bibinfo {author} {\bibfnamefont {C.~M.}\ \bibnamefont
  {Bishop}},\ }\href@noop {} {\emph {\bibinfo {title} {Pattern Recognition and
  Machine Learning (Information Science and Statistics)}}}\ (\bibinfo
  {publisher} {Springer-Verlag},\ \bibinfo {address} {Berlin, Heidelberg},\
  \bibinfo {year} {2006})\BibitemShut {NoStop}%
\bibitem [{\citenamefont {Mehta}\ \emph {et~al.}(2019)\citenamefont {Mehta},
  \citenamefont {Bukov}, \citenamefont {Wang}, \citenamefont {Day},
  \citenamefont {Richardson}, \citenamefont {Fisher},\ and\ \citenamefont
  {Schwab}}]{mehta_19}%
  \BibitemOpen
  \bibfield  {author} {\bibinfo {author} {\bibfnamefont {P.}~\bibnamefont
  {Mehta}}, \bibinfo {author} {\bibfnamefont {M.}~\bibnamefont {Bukov}},
  \bibinfo {author} {\bibfnamefont {C.-H.}\ \bibnamefont {Wang}}, \bibinfo
  {author} {\bibfnamefont {A.~G.}\ \bibnamefont {Day}}, \bibinfo {author}
  {\bibfnamefont {C.}~\bibnamefont {Richardson}}, \bibinfo {author}
  {\bibfnamefont {C.~K.}\ \bibnamefont {Fisher}},\ and\ \bibinfo {author}
  {\bibfnamefont {D.~J.}\ \bibnamefont {Schwab}},\ }\bibfield  {title}
  {\bibinfo {title} {A high-bias, low-variance introduction to machine learning
  for physicists},\ }\href
  {https://doi.org/https://doi.org/10.1016/j.physrep.2019.03.001} {\bibfield
  {journal} {\bibinfo  {journal} {Physics Reports}\ }\textbf {\bibinfo {volume}
  {810}},\ \bibinfo {pages} {1} (\bibinfo {year} {2019})},\ \bibinfo {note} {a
  high-bias, low-variance introduction to Machine Learning for
  physicists}\BibitemShut {NoStop}%
\bibitem [{\citenamefont {Efremkin}\ \emph {et~al.}(2021)\citenamefont
  {Efremkin}, \citenamefont {Barrat}, \citenamefont {Mossa},\ and\
  \citenamefont {Holzmann}}]{efremkin_21}%
  \BibitemOpen
  \bibfield  {author} {\bibinfo {author} {\bibfnamefont {V.}~\bibnamefont
  {Efremkin}}, \bibinfo {author} {\bibfnamefont {J.-L.}\ \bibnamefont
  {Barrat}}, \bibinfo {author} {\bibfnamefont {S.}~\bibnamefont {Mossa}},\ and\
  \bibinfo {author} {\bibfnamefont {M.}~\bibnamefont {Holzmann}},\ }\bibfield
  {title} {\bibinfo {title} {{Time correlation functions for quantum systems:
  Validating Bayesian approaches for harmonic oscillators and beyond}},\ }\href
  {https://doi.org/10.1063/5.0057279} {\bibfield  {journal} {\bibinfo
  {journal} {The Journal of Chemical Physics}\ }\textbf {\bibinfo {volume}
  {155}},\ \bibinfo {pages} {134108} (\bibinfo {year} {2021})}\BibitemShut
  {NoStop}%
\bibitem [{\citenamefont {Yang}\ and\ \citenamefont
  {Feiguin}(2021)}]{feiguin_21}%
  \BibitemOpen
  \bibfield  {author} {\bibinfo {author} {\bibfnamefont {L.}~\bibnamefont
  {Yang}}\ and\ \bibinfo {author} {\bibfnamefont {A.~E.}\ \bibnamefont
  {Feiguin}},\ }\bibfield  {title} {\bibinfo {title} {{From deconfined spinons
  to coherent magnons in an antiferromagnetic Heisenberg chain with long range
  interactions}},\ }\href {https://doi.org/10.21468/SciPostPhys.10.5.110}
  {\bibfield  {journal} {\bibinfo  {journal} {SciPost Phys.}\ }\textbf
  {\bibinfo {volume} {10}},\ \bibinfo {pages} {110} (\bibinfo {year}
  {2021})}\BibitemShut {NoStop}%
\bibitem [{\citenamefont {Yang}\ \emph {et~al.}(tion)\citenamefont {Yang},
  \citenamefont {Schumm},\ and\ \citenamefont {Sandvik}}]{yang_24}%
  \BibitemOpen
  \bibfield  {author} {\bibinfo {author} {\bibfnamefont {S.}~\bibnamefont
  {Yang}}, \bibinfo {author} {\bibfnamefont {G.}~\bibnamefont {Schumm}},\ and\
  \bibinfo {author} {\bibfnamefont {A.~W.}\ \bibnamefont {Sandvik}},\
  }\href@noop {} {\bibinfo {title} {Dynamic structure factor of spin-1/2 chains
  with long-range interactions}} (\bibinfo {year} {in preperation})\BibitemShut
  {NoStop}%
\bibitem [{\citenamefont {Shao}\ \emph {et~al.}(2017)\citenamefont {Shao},
  \citenamefont {Qin}, \citenamefont {Capponi}, \citenamefont {Chesi},
  \citenamefont {Meng},\ and\ \citenamefont {Sandvik}}]{sandvik_17}%
  \BibitemOpen
  \bibfield  {author} {\bibinfo {author} {\bibfnamefont {H.}~\bibnamefont
  {Shao}}, \bibinfo {author} {\bibfnamefont {Y.~Q.}\ \bibnamefont {Qin}},
  \bibinfo {author} {\bibfnamefont {S.}~\bibnamefont {Capponi}}, \bibinfo
  {author} {\bibfnamefont {S.}~\bibnamefont {Chesi}}, \bibinfo {author}
  {\bibfnamefont {Z.~Y.}\ \bibnamefont {Meng}},\ and\ \bibinfo {author}
  {\bibfnamefont {A.~W.}\ \bibnamefont {Sandvik}},\ }\bibfield  {title}
  {\bibinfo {title} {Nearly deconfined spinon excitations in the square-lattice
  spin-$1/2$ heisenberg antiferromagnet},\ }\href
  {https://doi.org/10.1103/PhysRevX.7.041072} {\bibfield  {journal} {\bibinfo
  {journal} {Phys. Rev. X}\ }\textbf {\bibinfo {volume} {7}},\ \bibinfo {pages}
  {041072} (\bibinfo {year} {2017})}\BibitemShut {NoStop}%
\bibitem [{\citenamefont {Canali}\ and\ \citenamefont
  {Wallin}(1993)}]{wallin_93}%
  \BibitemOpen
  \bibfield  {author} {\bibinfo {author} {\bibfnamefont {C.~M.}\ \bibnamefont
  {Canali}}\ and\ \bibinfo {author} {\bibfnamefont {M.}~\bibnamefont
  {Wallin}},\ }\bibfield  {title} {\bibinfo {title} {Spin-spin correlation
  functions for the square-lattice heisenberg antiferromagnet at zero
  temperature},\ }\href {https://doi.org/10.1103/PhysRevB.48.3264} {\bibfield
  {journal} {\bibinfo  {journal} {Phys. Rev. B}\ }\textbf {\bibinfo {volume}
  {48}},\ \bibinfo {pages} {3264} (\bibinfo {year} {1993})}\BibitemShut
  {NoStop}%
\bibitem [{\citenamefont {Igarashi}(1992)}]{igarashi_92}%
  \BibitemOpen
  \bibfield  {author} {\bibinfo {author} {\bibfnamefont {J.-i.}\ \bibnamefont
  {Igarashi}},\ }\bibfield  {title} {\bibinfo {title} {1/s expansion for
  thermodynamic quantities in a two-dimensional heisenberg antiferromagnet at
  zero temperature},\ }\href {https://doi.org/10.1103/PhysRevB.46.10763}
  {\bibfield  {journal} {\bibinfo  {journal} {Phys. Rev. B}\ }\textbf {\bibinfo
  {volume} {46}},\ \bibinfo {pages} {10763} (\bibinfo {year}
  {1992})}\BibitemShut {NoStop}%
\bibitem [{\citenamefont {Caux}\ and\ \citenamefont
  {Maillet}(2005)}]{maillet_05}%
  \BibitemOpen
  \bibfield  {author} {\bibinfo {author} {\bibfnamefont {J.-S.}\ \bibnamefont
  {Caux}}\ and\ \bibinfo {author} {\bibfnamefont {J.~M.}\ \bibnamefont
  {Maillet}},\ }\bibfield  {title} {\bibinfo {title} {Computation of dynamical
  correlation functions of heisenberg chains in a magnetic field},\ }\href
  {https://doi.org/10.1103/PhysRevLett.95.077201} {\bibfield  {journal}
  {\bibinfo  {journal} {Phys. Rev. Lett.}\ }\textbf {\bibinfo {volume} {95}},\
  \bibinfo {pages} {077201} (\bibinfo {year} {2005})}\BibitemShut {NoStop}%
\bibitem [{\citenamefont {Caux}\ \emph
  {et~al.}(2005{\natexlab{a}})\citenamefont {Caux}, \citenamefont {Hagemans},\
  and\ \citenamefont {Maillet}}]{caux_2005}%
  \BibitemOpen
  \bibfield  {author} {\bibinfo {author} {\bibfnamefont {J.-S.}\ \bibnamefont
  {Caux}}, \bibinfo {author} {\bibfnamefont {R.}~\bibnamefont {Hagemans}},\
  and\ \bibinfo {author} {\bibfnamefont {J.~M.}\ \bibnamefont {Maillet}},\
  }\bibfield  {title} {\bibinfo {title} {Computation of dynamical correlation
  functions of heisenberg chains: the gapless anisotropic regime},\ }\href
  {https://doi.org/10.1088/1742-5468/2005/09/P09003} {\bibfield  {journal}
  {\bibinfo  {journal} {Journal of Statistical Mechanics: Theory and
  Experiment}\ }\textbf {\bibinfo {volume} {2005}},\ \bibinfo {pages} {P09003}
  (\bibinfo {year} {2005}{\natexlab{a}})}\BibitemShut {NoStop}%
\bibitem [{\citenamefont {Pereira}\ \emph {et~al.}(2006)\citenamefont
  {Pereira}, \citenamefont {Sirker}, \citenamefont {Caux}, \citenamefont
  {Hagemans}, \citenamefont {Maillet}, \citenamefont {White},\ and\
  \citenamefont {Affleck}}]{affleck_06}%
  \BibitemOpen
  \bibfield  {author} {\bibinfo {author} {\bibfnamefont {R.~G.}\ \bibnamefont
  {Pereira}}, \bibinfo {author} {\bibfnamefont {J.}~\bibnamefont {Sirker}},
  \bibinfo {author} {\bibfnamefont {J.-S.}\ \bibnamefont {Caux}}, \bibinfo
  {author} {\bibfnamefont {R.}~\bibnamefont {Hagemans}}, \bibinfo {author}
  {\bibfnamefont {J.~M.}\ \bibnamefont {Maillet}}, \bibinfo {author}
  {\bibfnamefont {S.~R.}\ \bibnamefont {White}},\ and\ \bibinfo {author}
  {\bibfnamefont {I.}~\bibnamefont {Affleck}},\ }\bibfield  {title} {\bibinfo
  {title} {Dynamical spin structure factor for the anisotropic spin-$1/2$
  heisenberg chain},\ }\href {https://doi.org/10.1103/PhysRevLett.96.257202}
  {\bibfield  {journal} {\bibinfo  {journal} {Phys. Rev. Lett.}\ }\textbf
  {\bibinfo {volume} {96}},\ \bibinfo {pages} {257202} (\bibinfo {year}
  {2006})}\BibitemShut {NoStop}%
\bibitem [{\citenamefont {Sandvik}(2010)}]{sandvik_10}%
  \BibitemOpen
  \bibfield  {author} {\bibinfo {author} {\bibfnamefont {A.~W.}\ \bibnamefont
  {Sandvik}},\ }\bibfield  {title} {\bibinfo {title} {{Computational Studies of
  Quantum Spin Systems}},\ }\href {https://doi.org/10.1063/1.3518900}
  {\bibfield  {journal} {\bibinfo  {journal} {AIP Conference Proceedings}\
  }\textbf {\bibinfo {volume} {1297}},\ \bibinfo {pages} {135} (\bibinfo {year}
  {2010})}\BibitemShut {NoStop}%
\bibitem [{cau()}]{caux_BA}%
  \BibitemOpen
  \href@noop {} {}\bibinfo {note} {BA results from Ref.~\cite{caux_2005}
  provided by J.-S. Caux (private communication).}\BibitemShut {Stop}%
\bibitem [{\citenamefont {Caux}\ \emph
  {et~al.}(2005{\natexlab{b}})\citenamefont {Caux}, \citenamefont {Hagemans},\
  and\ \citenamefont {Maillet}}]{caux_05}%
  \BibitemOpen
  \bibfield  {author} {\bibinfo {author} {\bibfnamefont {J.-S.}\ \bibnamefont
  {Caux}}, \bibinfo {author} {\bibfnamefont {R.}~\bibnamefont {Hagemans}},\
  and\ \bibinfo {author} {\bibfnamefont {J.~M.}\ \bibnamefont {Maillet}},\
  }\bibfield  {title} {\bibinfo {title} {Computation of dynamical correlation
  functions of heisenberg chains: the gapless anisotropic regime},\ }\href
  {https://doi.org/10.1088/1742-5468/2005/09/P09003} {\bibfield  {journal}
  {\bibinfo  {journal} {Journal of Statistical Mechanics: Theory and
  Experiment}\ }\textbf {\bibinfo {volume} {2005}},\ \bibinfo {pages} {P09003}
  (\bibinfo {year} {2005}{\natexlab{b}})}\BibitemShut {NoStop}%
\bibitem [{\citenamefont {Wang}\ and\ \citenamefont {Lin}(2019)}]{wang_19}%
  \BibitemOpen
  \bibfield  {author} {\bibinfo {author} {\bibfnamefont {L.}~\bibnamefont
  {Wang}}\ and\ \bibinfo {author} {\bibfnamefont {H.-Q.}\ \bibnamefont {Lin}},\
  }\href@noop {} {\bibinfo {title} {Dynamic structure factor from real time
  evolution and exact correction vectors with matrix product states}} (\bibinfo
  {year} {2019}),\ \Eprint {https://arxiv.org/abs/1901.07751} {arXiv:1901.07751
  [cond-mat.str-el]} \BibitemShut {NoStop}%
\bibitem [{\citenamefont {Xie}\ \emph {et~al.}(2018)\citenamefont {Xie},
  \citenamefont {Huang}, \citenamefont {Han}, \citenamefont {Yan},
  \citenamefont {Zhao}, \citenamefont {Xie}, \citenamefont {Liao},\ and\
  \citenamefont {Xiang}}]{xie_18}%
  \BibitemOpen
  \bibfield  {author} {\bibinfo {author} {\bibfnamefont {H.~D.}\ \bibnamefont
  {Xie}}, \bibinfo {author} {\bibfnamefont {R.~Z.}\ \bibnamefont {Huang}},
  \bibinfo {author} {\bibfnamefont {X.~J.}\ \bibnamefont {Han}}, \bibinfo
  {author} {\bibfnamefont {X.}~\bibnamefont {Yan}}, \bibinfo {author}
  {\bibfnamefont {H.~H.}\ \bibnamefont {Zhao}}, \bibinfo {author}
  {\bibfnamefont {Z.~Y.}\ \bibnamefont {Xie}}, \bibinfo {author} {\bibfnamefont
  {H.~J.}\ \bibnamefont {Liao}},\ and\ \bibinfo {author} {\bibfnamefont
  {T.}~\bibnamefont {Xiang}},\ }\bibfield  {title} {\bibinfo {title}
  {Reorthonormalization of chebyshev matrix product states for dynamical
  correlation functions},\ }\href {https://doi.org/10.1103/PhysRevB.97.075111}
  {\bibfield  {journal} {\bibinfo  {journal} {Phys. Rev. B}\ }\textbf {\bibinfo
  {volume} {97}},\ \bibinfo {pages} {075111} (\bibinfo {year}
  {2018})}\BibitemShut {NoStop}%
\bibitem [{\citenamefont {Yusuf}\ \emph {et~al.}(2004)\citenamefont {Yusuf},
  \citenamefont {Joshi},\ and\ \citenamefont {Yang}}]{yusuf_04}%
  \BibitemOpen
  \bibfield  {author} {\bibinfo {author} {\bibfnamefont {E.}~\bibnamefont
  {Yusuf}}, \bibinfo {author} {\bibfnamefont {A.}~\bibnamefont {Joshi}},\ and\
  \bibinfo {author} {\bibfnamefont {K.}~\bibnamefont {Yang}},\ }\bibfield
  {title} {\bibinfo {title} {Spin waves in antiferromagnetic spin chains with
  long-range interactions},\ }\href
  {https://doi.org/10.1103/PhysRevB.69.144412} {\bibfield  {journal} {\bibinfo
  {journal} {Phys. Rev. B}\ }\textbf {\bibinfo {volume} {69}},\ \bibinfo
  {pages} {144412} (\bibinfo {year} {2004})}\BibitemShut {NoStop}%
\bibitem [{\citenamefont {Laflorencie}\ \emph
  {et~al.}(2005{\natexlab{a}})\citenamefont {Laflorencie}, \citenamefont
  {Affleck},\ and\ \citenamefont {Berciu}}]{laflorencie_05}%
  \BibitemOpen
  \bibfield  {author} {\bibinfo {author} {\bibfnamefont {N.}~\bibnamefont
  {Laflorencie}}, \bibinfo {author} {\bibfnamefont {I.}~\bibnamefont
  {Affleck}},\ and\ \bibinfo {author} {\bibfnamefont {M.}~\bibnamefont
  {Berciu}},\ }\bibfield  {title} {\bibinfo {title} {Critical phenomena and
  quantum phase transition in long range heisenberg antiferromagnetic chains},\
  }\href {https://doi.org/10.1088/1742-5468/2005/12/P12001} {\bibfield
  {journal} {\bibinfo  {journal} {Journal of Statistical Mechanics: Theory and
  Experiment}\ }\textbf {\bibinfo {volume} {2005}},\ \bibinfo {pages} {P12001}
  (\bibinfo {year} {2005}{\natexlab{a}})}\BibitemShut {NoStop}%
\bibitem [{\citenamefont {Laflorencie}\ \emph
  {et~al.}(2005{\natexlab{b}})\citenamefont {Laflorencie}, \citenamefont
  {Affleck},\ and\ \citenamefont {Berciu}}]{laflorencie_2005}%
  \BibitemOpen
  \bibfield  {author} {\bibinfo {author} {\bibfnamefont {N.}~\bibnamefont
  {Laflorencie}}, \bibinfo {author} {\bibfnamefont {I.}~\bibnamefont
  {Affleck}},\ and\ \bibinfo {author} {\bibfnamefont {M.}~\bibnamefont
  {Berciu}},\ }\bibfield  {title} {\bibinfo {title} {Critical phenomena and
  quantum phase transition in long range heisenberg antiferromagnetic chains},\
  }\href {https://doi.org/10.1088/1742-5468/2005/12/P12001} {\bibfield
  {journal} {\bibinfo  {journal} {Journal of Statistical Mechanics: Theory and
  Experiment}\ }\textbf {\bibinfo {volume} {2005}},\ \bibinfo {pages} {P12001}
  (\bibinfo {year} {2005}{\natexlab{b}})}\BibitemShut {NoStop}%
\bibitem [{\citenamefont {Quan-Sheng}\ \emph {et~al.}(2013)\citenamefont
  {Quan-Sheng}, \citenamefont {Yi-Lin}, \citenamefont {Zhong},\ and\
  \citenamefont {Xi}}]{wu_13}%
  \BibitemOpen
  \bibfield  {author} {\bibinfo {author} {\bibfnamefont {W.}~\bibnamefont
  {Quan-Sheng}}, \bibinfo {author} {\bibfnamefont {W.}~\bibnamefont {Yi-Lin}},
  \bibinfo {author} {\bibfnamefont {F.}~\bibnamefont {Zhong}},\ and\ \bibinfo
  {author} {\bibfnamefont {D.}~\bibnamefont {Xi}},\ }\bibfield  {title}
  {\bibinfo {title} {Acceleration of the stochastic analytic continuation
  method via an orthogonal polynomial representation of the spectral
  function},\ }\href {https://doi.org/10.1088/0256-307X/30/9/090201} {\bibfield
   {journal} {\bibinfo  {journal} {Chinese Physics Letters}\ }\textbf {\bibinfo
  {volume} {30}},\ \bibinfo {pages} {090201} (\bibinfo {year}
  {2013})}\BibitemShut {NoStop}%
\end{thebibliography}%

\end{document}